\documentclass[superscriptaddress,twocolumn,pre]{revtex4-1} 
\usepackage{epsfig}
\usepackage{graphicx}  
\usepackage{enumerate}
\usepackage{bm}
\usepackage{natbib}
\usepackage{amssymb,amsfonts,amsmath}
\usepackage{hyperref}

\newcommand{\bra}[1]{\left(#1\right)}
\newcommand{\Bra}[1]{\left[#1\right]}

\begin{document}


\title{Catalytic membrane reactor model as a laboratory for pattern emergence in reaction-diffusion-advection media}


\author{Arik Yochelis} \email{yochelis@bgu.ac.il}
\affiliation{Department of Solar Energy and Environmental Physics, Swiss Institute for Dryland Environmental and Energy Research, Blaustein Institutes for Desert Research, Ben-Gurion University of the Negev, Sede Boqer Campus, Midreshet Ben-Gurion 8499000, Israel} 
\affiliation{Department of Physics, Ben-Gurion University of the Negev, Be'er Sheva 8410501, Israel}

\received{\today}
\begin{abstract}
Reaction-diffusion-advection media on semi-infinite domains are important in chemical, biological and ecological applications, yet remain a challenge for pattern formation theory. To demonstrate the rich emergence of nonlinear traveling waves and stationary periodic states, we review results obtained using a membrane reactor as a case model. Such solutions coexist in overlapping parameter regimes and their temporal stability is determined by the boundary conditions (periodic vs. mixed) which either preserve or destroy the translational symmetry, i.e., selection mechanisms under realistic Danckwerts boundary conditions. A brief outlook is given at the end.
\end{abstract}

\maketitle

\section{Introduction}

Reaction diffusion (RD) models are known to exhibit universal self-organized patterns, such as spiral and solitary waves, standing-wave labyrinths, and oscillating spots~\cite{CrHo:93,MPC:97,kapral2012chemical}. As such, the seminal work of Turing~\cite{Tur:52} and its extensions~\cite{keener1998mathematical,Mu:02,Pis:06,Meron2015}) became central in the theory of pattern formation~\cite{Hoyle:06} and led to myraid of fundamental insights into to pattern selection mechanisms that show up across many fields of applied science. Yet, in cases where the pattern forming instabilities are sub-critical, i.e., \textit{nonlinear} instabilities, several distinct spatially nonuniform states may be found to coexist under the same conditions. The selection mechanisms in these situations are still intriguing open problems~\cite{Kn:02}.

A fundamental and realistic extension of RD media is the inclusion of a unidirectional reactant supply and product removal, i.e., transport by advection at different rates. Such a class of problems is often referred to as a reaction-diffusion-advection (RDA) medium. Mathematically, the advective fields destroy the translational symmetry of the system and thereby, distinguish between absolute instabilities (like in RD case) and convective instabilities~\cite{HuMo:90}, which belong to the class of nonlinear instabilities~\cite{Chom:92}. RDA systems exhibit not only a wide range of traveling and solitary waves but also stationary nonuniform patterns under certain BCs~\cite{yakhnin1995convective,KhPi:95,kosek1995splitting,Sh:97,KMDB:97,SMS:98,ABMDB:99,NRS:00,SM:00,BKMS:00,SMM:00,SMM:01,Bam:01,NeSh:02,KM:02,ShNe:03,Sat:03,NeSh:03,MM:05,MSM:06,ZMK:06,FSN:07,YNBM:07,VMS:08}. This review focuses on the \textit{nonlinear} selection mechanisms in the presence of multiple co-existing solutions, which include the effects imposed by BCs. Understanding these mechanisms is an intriguing problem in pattern formation theory.

Systems involving RDA processes may arise in a broad class of applied sciences, including tubular reactors~\cite{sheintuch1996spatiotemporal}, axial segmentation in
vertebrates~\cite{KMSH:02}, biochemical oscillations in the amoeboid organism Physarum~\cite{YNBM:07}, autocatalytic reactions on a
rotating disk~\cite{KhPi:95}, vegetation patterns~\cite{BDLR:09}, and thus have been studied both analytically and numerically~\cite{RoMe:93,KhPi:95,KMDB:97,Sh:97,SMS:98,Klaus:99,ABMDB:99,KM:99,NRS:00,SM:00,BKMS:00,SMM:00,NNRS:00,SMM:01,Bam:01,NeSh:02,KM:02,ShNe:03,Sat:03,NeSh:03,MM:05,MSM:06,ZMK:06,FSN:07,YNBM:07,VMS:08,ABMDB:99,couairon1999primary}.

The theoretical effort to date has mainly been devoted to the region in the proximity of the instability. This approach leaves many questions unresolved, such as the effect of nonlinear instabilities and boundary conditions on spatiotemporal dynamics, such as discussed in~\cite{deissler1985noise,Chom:92,MuTv:95,CoCh:97,TPK:98,NNRS:00,MM:05,FSN:07}. In pursuit of the pattern selection mechanism at work, we set out to review the methods of spatial dynamics and how they serve as a powerful framework for analyzing problems of this type: bifurcation analysis of nonuniform states coupled with numerical continuation and temporal eigenvalue computations to identify the stability of the obtained solutions. The objective is to provide a brief description of the distinct from RD, nonlinear pattern selection mechanisms of traveling waves (TW), pulses, and stationary periodic (SP) patterns under a semi-infinite one dimensional spatial domain (1D), with periodic or Danckwerts-type BCs. Further details may be found in~\cite{yochelis2009principal,yochelis2009towards,yochelis2010drifting}. 

\section{Catalytic membrane reactor by Sheintuch and Nekhamkina}

We demonstrate the following results through a model of a pseudo-homogeneous catalytic membrane reactor~\cite{sheintuch1996spatiotemporal} in which a single first order exothermic reaction occurs, or a simple flow reactor with two consecutive reactions $A\longrightarrow B \longrightarrow C$, where the first reaction proceeds at a constant rate;  similar model has been also used for a one-dimensional tubular cross-flow reactor describing $A \to B + \mathrm{heat}$~\cite{yakhnin1994differential,yakhnin1994differential_b}. The reactants in a reactor are supplied along the systems to avoid temperature runaway or poor selectivity that may be associated with feed at one point. The mass and energy balances can be written in dimensionless form~\cite{ShNe:99,NRS:00}
\begin{subequations}\label{eq:PDE}
	\begin{eqnarray}
	\frac{\partial u}{\partial t}+ \frac{\partial u}{\partial x} &=& f\bra{u,v}-u, \\
	Le\frac{\partial v}{\partial t}+ \frac{\partial v}{\partial x}&=&B f\bra{u,v}-\alpha v +\frac{1}{Pe}\frac{\partial^2 v}{\partial x^2},
	\label{eq:PDE_adv}
	\end{eqnarray}
	where
	\begin{eqnarray}\label{eq:fun}
	\quad f\bra{u,v} \equiv Da (1-u)\exp\Bra{\dfrac{\gamma v}{\gamma+v}},
	\end{eqnarray}
\end{subequations}
{describes a simple a first order exothermic reaction of Arrhenius kinetics and is being used for many reactor design problems, for understanding instabilities, explosions and cool flames~\cite{sheintuch1996spatiotemporal,Pis:06}}. In~(\ref{eq:PDE}) $u(x,t)$ stands for conversion ($u = 1$ implies zero reactant concentration) and can be viewed as a fast inhibitor while $v(x,t)$ is the temperature or a slow activator in the context of RD systems, $Da$ is the Damk\"{o}hler number describing the rate of an activated reaction (Arrhenius kinetics~\cite{URP:74}), $Le$ is the Lewis number that is associated with the ratio of solid- to fluid-phase heat capacities (assumed to be large), and $Pe$ is the P\'{e}clet number that is associated with the ratio of convective to conductive enthalpy fluxes (assumed to be large to support steep gradients). 

Cross-flow systems are semi-infinite, so that the BCs correspond to mixed at the inlet
\begin{subequations}\label{eq:BC}
	\begin{equation}
	a_uu+b_u\frac{\partial u}{\partial x}{\bigg |}_{x=0}=g_u,\quad a_vv+b_v\frac{\partial v}{\partial x}{\bigg |}_{x=0}=g_v, 
	\end{equation}
	and no-flux at the outlet
	\begin{equation}
	\frac{\partial v}{\partial x}{\bigg |}_{x=L}=0, 
	\end{equation}
\end{subequations}
	where $L$ is the physical domain size, and $a_{u,v}$, $b_{u,v}$ and $g_{u,v}$ are real constants. The realistic BC of Danckwerts type~\cite{froment2011chemical} correspond to
	\begin{eqnarray}\label{eq:BC_Dank}
	\begin{array}{l}
	\left( {\begin{array}{*{20}c}
		{a_u }  \\
		{b_u }  \\
		{g_u }  \\
		\end{array}} \right) = \left( {\begin{array}{*{20}c}
		1  \\
		0  \\
		0  \\
		\end{array}} \right),\quad	
	\left( {\begin{array}{*{20}c}
		{a_v }  \\
		{b_v }  \\
		{g_v }  \\
		\end{array}} \right) = \left( {\begin{array}{*{20}c}
		1  \\
		{ -Pe^{-1}}  \\
		0  \\
		\end{array}} \right) \\ 
	\end{array}.
	\end{eqnarray}
Notably, equations similar to~\eqref{eq:PDE} also describe the high-switching asymptote of a loop reactor, where the feed is periodically switched between several units~\cite{ShNe:05}, but the BC in this case are periodic.

\begin{figure}[tp]
	\includegraphics[width=0.4\textwidth]{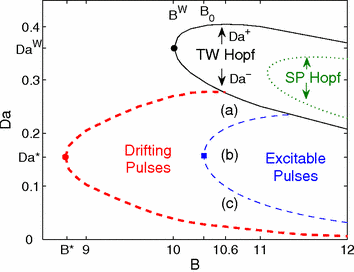}
	\caption{Parameter space spanned by $(B,Da)$ showing the regions of excitable and drifting pulses together with the finite wavenumber instability of counter-propagating traveling waves (TW), $Da^\pm$, and the the criterion for stationary periodic patterns, SP Hopf (dotted line). Reprinted from~\cite{yochelis2010drifting}.} \label{fig:1}
\end{figure}

Direct numerical integrations of~\eqref{eq:PDE} with~\eqref{eq:BC_Dank} show that traveling waves, pulses, and stationary periodic patterns are persistent solutions of the system, as summarized in the parameter space that is presented in Fig.~\ref{fig:1}). The rest of this review is devoted to examination of the pattern selection mechanisms that operate in systems of the type described by~\eqref{eq:PDE}.
\begin{figure*}[htb]
	\includegraphics[width=0.65\textwidth]{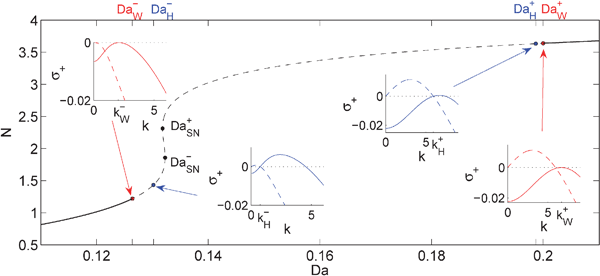}
	\caption{Bifurcation diagram for uniform $(u_0,v_0)$ solutions as a function of $Da$, underlying bistability within $Da_{SN}^-\leq Da\leq Da_{SN}^+$; temporal stability to uniform perturbations is identified by solid lines. The right- and the left-most insets indicate the onset of finite wavenumber Hopf bifurcations, $Da_W^\pm$, at which $Re[\sigma_+(k^\pm_W)]=0$, while the middle dispersion relations mark the condition $Re[\sigma_+(k^\pm_H)]=Im[\sigma_+(k^\pm_H)]=0$ for spatially periodic patterns obtained at $Da_H^\pm$; solid and dashed lines mark $Re[\sigma_+(k)]$ and $Im[\sigma_+(k)]$, respectively. Reprinted from~\cite{yochelis2009towards}.} \label{fig:2}
\end{figure*}

\section{Linear theory}

It is useful to start with an infinite domain in which traveling waves emerge from a finite wavenumber Hopf bifurcation about a uniform steady state, $(u,v)=(u_0,v_0)$, which result from
\[
Da-\frac{u_0}{1-u_0}\exp\Bra{-\dfrac{\gamma u_0}{\gamma \alpha/B+u_0}}=0,\quad v_0\equiv B u_0/\alpha.
\]
These uniform solutions are organized in a cusp bifurcation, i.e., they exhibit mono- or bi-stability, where the coexistence regime, under variation of $Da$, lies in between two saddle nodes $Da^+_{SN}\leq Da\leq Da^-_{SN}$~\cite{URP:74}, as depicted in Fig.~\ref{fig:2}.

\subsection{Dispersion relation and instability to traveling waves}
Linear stability analysis to periodic perturbations about the uniform state $(u_0,v_0)$ is approached by examining
\begin{equation}\label{eq:lin}
\left( {\begin{array}{c}
	u  \\
	v  \\
	\end{array}} \right) - \left( {\begin{array}{c}
	{u_0 }  \\
	{v_0 }  \\
	\end{array}} \right) \propto  e^{\sigma t+ikx}+c.c.+h.o.t.,
\end{equation}
where $\sigma$ is the (complex) perturbation growth rate, $k>0$ is the wavenumber, $c.c.$ denotes a complex conjugate, and $h.o.t.$ stand for high order terms. The standard calculation yields two dispersion relations $\sigma_\pm(k)$, of which only $\sigma_+(k)$, is relevant, as $Re[\sigma_+(k^\pm_W)]=0$ at $Da=Da^\pm_W$ indicates the onset of a finite wavenumber Hopf instability while $Re[\sigma_-(k)]<0$ for all $k$, with $k^+_W$ and $k^-_W$ respectively denoting the critical wavenumbers of the upper and the lower branches of $(u_0,v_0)$, see Figure~\ref{fig:2}. Notably, the speed and direction of the TW is dictated by $Re[\sigma_+(k^\pm_W)]$ and is found to be negative at both onsets. However, this analysis is incomplete, as in RDA systems the type of instability can be either convective or absolute~\cite{HuMo:90,Chom:92,CoCh:97}, but since the interest here is in pattern formation far from instability onsets, the reader is referred to~\cite{NNRS:00}. In the absence of differential flow, i.e., in RD media, the finite wavenumber Hopf bifurcation is encountered in a \textit{three}-component system with at least two diffusing fields~\cite{Yo:08,anma2012unstable,hata2014sufficient}, giving rise to both traveling and standing waves~\cite{knobloch1986oscillatory}. Here, the advective terms in~(\ref{eq:PDE_adv}) break the spatial reflection symmetry of right-left propagating waves so that only one family is selected. This breaking of symmetry also precludes the emergence of standing waves. 

\begin{figure*}[htb]
	\includegraphics[width=0.9\textwidth]{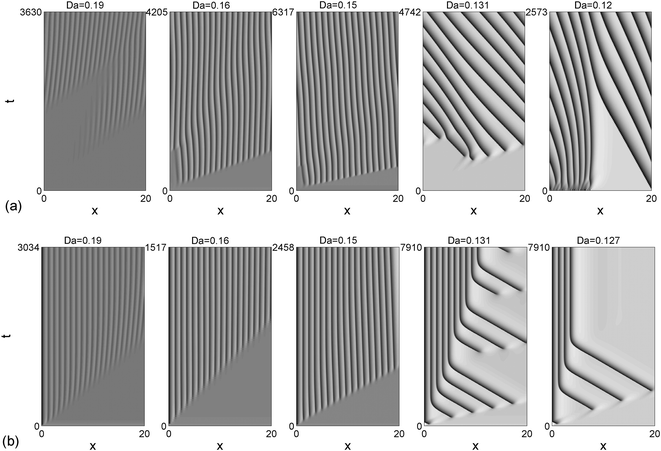}
	\caption{Direct numerical integration of (\ref{eq:PDE}) showing space-time plots for different values of $Da$ and with (a) periodic or (b) Danckwerts boundary conditions, where dark color indicates larger $v$ field values. Reprinted from~\cite{yochelis2009towards}.} \label{fig:3}
\end{figure*}

\begin{figure}[htb]
	\includegraphics[width=0.35\textwidth]{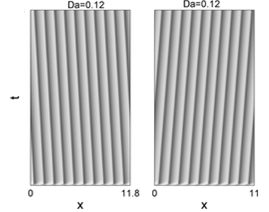}
	\caption{Direct numerical integration of (\ref{eq:PDE}) with periodic boundary conditions represented as space-time plot, where dark color indicates larger $v$ values. Reprinted from~\cite{yochelis2009towards}.} \label{fig:4}
\end{figure}

Although linear theory predicts emergence of traveling waves, direct numerical simulations of Eq.~\ref{eq:PDE} show several intriguing features of the pattern selection mechanisms:
\begin{description}
	\item [Periodic BC] Figure~\ref{fig:3}(a) shows that for $Da\lesssim Da^+_W$, the instability is of convective type and small-amplitude traveling waves propagating towards the outlet emerge. As $Da$ is decreased toward $Da^-_W$ the instability becomes absolute but the propagation direction of the TW changes toward the inlet. When $Da$ is decreased further the period of the TW increases and large amplitude TW persist also below $Da^-_W$, even though the uniform solution $(u_0,v_0)$ is linearly stable. Surprisingly, by controlling the initial perturbation and domain size, it is possible to obtain counter-propagating outlet-bound TW with a much shorter period that coexist with the inlet-bound waves at $Da<Da^-_W$, as shown in Fig.~\ref{fig:4}.
	\item [Danckwerts BC] Figure~\ref{fig:3}(b) shows that for $Da^-_W<Da<Da^+_W$ the asymptotic solutions are stationary periodic, while TW are transient although they exhibit the same features as described above with periodic BC. However, the period of the transient TW for $Da<Da^-_W$ is much larger than the period of TW under periodic BC. 
\end{description} 
Consequently, to understand the emergence of the above nonlinear patterns, and additionally other possible solutions, it is useful to exploit the spatial dynamics framework which, when used along with numerical continuation {methods}, allows efficient mapping of both stable and unstable solutions from which one can identify the pattern selection mechanisms.
\begin{figure*}[htb]
	\includegraphics[width=0.7\textwidth]{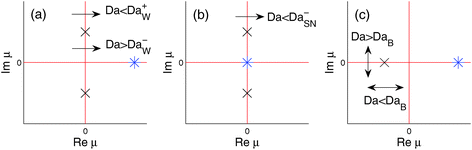}
	\caption{Schematic representation of the spatial eigenvalues about $(u_0,v_0)$, a complex conjugated pair ($\times$) and a real ($\ast$), at distinct bifurcation onsets: (a) Hopf at $Da=Da_W^\pm$, (b) saddle-node/Hopf at $Da=Da_{SN}^-$, and (c) Belyakov at $Da=Da_B$. The arrows indicate the motion of the complex eigenvalue pair as $Da$ is varied. Reprinted from~\cite{yochelis2009towards}.} \label{fig:5}
\end{figure*}
\begin{figure}[htb]
	\includegraphics[width=0.4\textwidth]{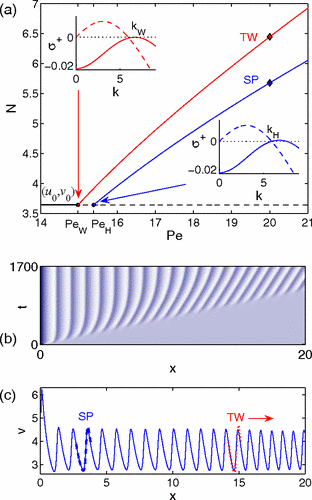}
	\caption{(a) Bifurcation diagram for spatially homogeneous ($u_0 ,v_0$) steady states [solid (stable) and dashed (unstable) dark lines] and distinct spatially periodic (light lines) solutions. The nonuniform solutions were numerically computed using Eq.~\ref{eq:ODE} with $Pe$ as a control parameter and on periodic domains. Traveling waves (TW) emerge from a finite wavenumber Hopf bifurcation at $Pe=Pe_W$. Stationary periodic (SP) solutions emerge from $Pe=Pe_H$. The insets represent the real (solid line) and	the imaginary (dashed line) parts of the respective dispersion relations using~\eqref{eq:PDE}. (b) Space-time plot where dark color indicates larger $v$ field values; Eq.~\ref{eq:PDE} was integrated starting from $(u(x),v(x))=(u_0,v_0)$ with Danckwerts boundary conditions. (c) A Spatial profile of $v(x)$ at specific time, where the dashed (dotted) line indicates the corresponding single period profile of SP (TW) state, obtained via the continuation method [see ($\blacklozenge$) symbols in (a)]. Reprinted from~\cite{yochelis2009principal}.} \label{fig:6super}
\end{figure}

\subsection{Spatial dynamics}
Spatial dynamics is a powerful methodology that allows exploiting the tools developed for ordinary differential equations (ODE) for analysis of nonuniform states in the spatially extended contexts to reveal the coexisting solutions in the parameter space of the problem; the stability properties of these solutions are obtained in the next stage by solving a temporal eigenvalue problem. This theoretical approach was shown to be useful for uncovering a number of complex nonlinear mechanisms, such as pattern formation in the presence of \textit{homoclinic snaking}, both in dissipative~\cite{burke2007homoclinic,YTDG:08,dawes2008localized,BYK:08,kozyreff2009influence,yochelis2015origin} and variational~\cite{thiele2013localized,gavish2017spatially} model equations.

For propagating solutions such as those observed in RDA, it is useful to consider~(\ref{eq:PDE}) in a co-moving frame, $\xi=x-ct$, where $c$ is the group velocity with the sign obtained by the dispersion relation at the onset, $c^\pm_W=Im[\sigma(k_W^\pm)]/k_W^\pm<0$. After the transformation 
\[ \partial_t \to \partial_t -c\partial_\xi, \quad \partial_x \to \partial_\xi, \] 
the time independent version of~\eqref{eq:PDE} reads, in the first order ODE form, as
\begin{subequations}\label{eq:ODE}
\begin{eqnarray}
\frac{\mathrm{d} u}{\mathrm{d} \xi}&=&\frac{1}{1-c} \Bra{f(u,v)-u}, \\
\frac{\mathrm{d} v}{\mathrm{d} \xi}&=&w \,, \\
\frac{\mathrm{d} w}{\mathrm{d} \xi}&=& Pe\Bra{\bra{1-c\, Le}w-B f(u,v)+\alpha v} \,.
\end{eqnarray}
\end{subequations}
Analysis of~\eqref{eq:ODE} also involves, as the first step, a linear analysis about the uniform states:
\begin{equation}\label{eq:eig}
\left( {\begin{array}{c}
	u  \\
	v  \\
	w
	\end{array}} \right) - \left( {\begin{array}{c}
	{u_0 }  \\
	{v_0 }  \\
	{0}
	\end{array}} \right) \propto  e^{\mu \xi}+h.o.t..
\end{equation}

Knowledge of the spatial eigenvalues reveals information about the onsets and characteristics of both propagating (with $c\neq 0$) and stationary (with $c=0$) nonuniform states. For example, the finite wavenumber Hopf instabilities that have been identified earlier at $Da=Da_W^\pm$ with $c=c_W^\pm$ correspond in (\ref{eq:ODE}) to Hopf bifurcations but in space with $c=c^\pm_W$, so that the configuration of the three spatial eigenvalues lies in the complex eigenvalue plane as schematically represented in Fig.~\ref{fig:5}(a). The splitting of the complex pair ($\times$) is: for $TW^+$, a purely imaginary pair becomes complex as $Da<Da_W^+$, while for $TW^-$, a purely imaginary pair becomes complex as $Da>Da_W^-$. In both cases, the real part of the pair is smaller than the third real eigenvalue ($\ast$), so that in regions $Da>Da_W^+$ and $Da<Da_W^-$ the linearization about the fixed point corresponds to a saddle-focus. Yet, only in the subcritical $Da<Da_W^-$ case does this lead to a homoclinic connection~\cite{Yo:08}. Other examples belong to the (codimension-two) saddle-node/Hopf bifurcation [Fig.~\ref{fig:5}(b)] at which a pure imaginary pair eigenvalues (of the Hopf type) coexists with a zero eigenvalue [of a fold of the uniform state ($u_0,v_0$)] and the so-called Belyakov point (at which a saddle focus of the linearized fixed point becomes a saddle) that represents the collision of the complex eigenvalue pair ($\times$) on the real axis~\cite{Bel:74,Bel:80}, at $Da\equiv Da_B$ and $c=0$, where for $Da>Da_B$ there is a complex pair and for $Da<Da_B$ the splitting is on the real axis so that all eigenvalues are real [Fig.~\ref{fig:5}(c)]. As will be shown next, these bifurcations will help to explain some of behaviors obtained via direct numerical integration.

\section{Time dependent Nonlinear solutions}

Since $Da^+_W$ is a super-critical bifurcation (i.e., an instability small amplitude $TW^+$), weakly nonlinear analysis in the form of a complex Ginzburg-Landau equation was used to understand the emergence of both TW and the SP solutions~\cite{NNRS:00}. Figure~\ref{fig:6super} demonstrates the respective computation using the spatial dynamics method and shows agreement with direct numerical integration. Yet, as in RD systems, this analysis cannot capture features that emerge at large distances from the onset, i.e., velocity changes of TW and the rich variety of patterns in the sub-critical regime of $Da^-_W$. Thus, the investigation of distinct solutions that emerge from the spatial bifurcations with either $c^\pm_W\neq 0$ (TW) or $c^\pm_W=0$ (SP) can be more efficiently advanced via numerical continuation {methods}, where $c$ is obtained by a nonlinear eigenvalue problem on periodic domains. Temporal stability of such solutions is computed via a standard numerical eigenvalue method using the time-dependent version of~\eqref{eq:ODE} in the co-moving frame and also by checking large domains $L=n\lambda$ for secondary instabilities, where $L$ is the domain size, $\lambda=2\pi/k$ is a single period of TW or SP, and $n$ is an integer. For clarity, the bifurcation diagrams are plotted in terms of a norm 
: {

\[
	N=\sqrt{\frac{1}{\lambda}\int_0^\lambda \Bra{\sum {y_i^2  + y\prime_i^2}}\mathrm{d}\xi},
\]	
where $y_i$ account for equation variables (here $u,v,w$) and $y\prime_i$ denote derivatives with respect to the argument (here $\xi$)}.

\begin{figure*}[htb]
	\includegraphics[width=0.65\textwidth]{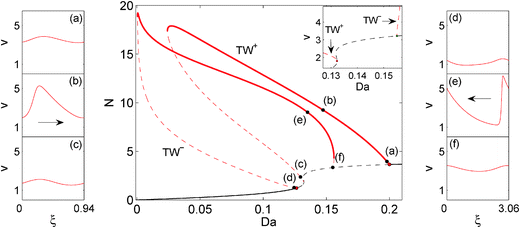}
	\caption{Middle panel: Bifurcation diagram for spatially homogeneous solutions, $(u_0,v_0,0)$, and traveling waves $TW^\pm$, as a function of $Da$. Eq.~(\ref{eq:ODE}) was integrated on periodic domains, where solid lines indicate temporal stability in periodic domains $L=\lambda_W^\pm$, in the context of Eq.~(\ref{eq:PDE}). The inset shows the respective oscillatory $TW^\pm$ branches in terms of the maximum value of $v$ in vicinity of termination points. Left panel (a-c) and right panel (d-f) correspond to profiles at $Da$ values as marked in the middle panel; the arrows in (b) and (e) mark the propagation direction in the context of Eq.~(\ref{eq:PDE}). Reprinted from~\cite{yochelis2009towards}.} \label{fig:6}
\end{figure*}
\begin{figure*}[tp]
	\includegraphics[width=0.8\textwidth]{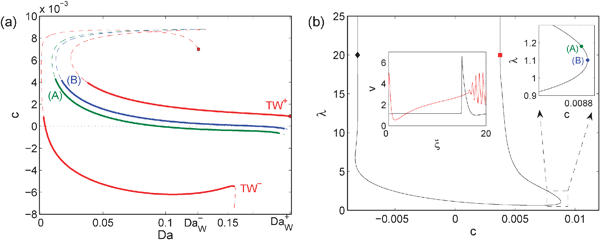}
	\caption{(a) Branches and stability of traveling wave solutions in $(Da,c)$ parameter space while periodic domain size $L=n\lambda \gg \lambda^\pm_W$ (along each branch) is fixed. (b) Nonlinear dispersion relation, $\lambda$ vs. $c$, slightly below the $Da=Da^-_W$ onset. The right inset magnifies the turning point where the points (A) and (B) are associated with secondary TW in (a). The left inset depicts profiles of spatially localized solutions of homoclinic type [connection to a fixed point ($\blacklozenge$) and to a periodic orbit ($\blacksquare$)]. Reprinted from~\cite{yochelis2009towards}.} \label{fig:7}
\end{figure*}

\subsection{Counter propagating traveling waves}

Primary $TW^\pm$ solutions arise from the two onsets $Da^\pm_W$ as periodic bifurcating orbits (Fig.~\ref{fig:6}), with respective fixed periods of $\lambda_W^\pm$.

As we have already shown, the $TW^+$ solutions bifurcate super-critically from $Da=Da_W^+$ and are right-propagating waves in the context of~(\ref{eq:PDE}) (see Fig.~\ref{fig:6}). Continuation of these oscillatory states to lower $Da$ values shows that stable states with $c>0$ and fixed spatial period, $\lambda=\lambda_W^+$, persist up to a fold (saddle-node) and then become unstable. After the fold, $TW^+$ terminate at $Da=Da_{SN}^-$ which is the (codimension-two) saddle-node/Hopf bifurcation [Fig.~\ref{fig:5}(b)]. The branch and typical $TW^+$ profiles are shown in Fig.~\ref{fig:6}. 

On the other hand, $TW^-$ solutions (with  $\lambda=\lambda_W^-$), that bifurcate from $Da=Da_W^-$, advance toward the linearly stable region, $Da<Da_W^-$, as unstable orbits, thereby forming a sub-critical bifurcation, as shown in Fig.~\ref{fig:6}. The branch folds around $Da \simeq 0$ and gains stability before extending itself to large $Da$ values, see Fig.~\ref{fig:6}. After an additional fold (at the rightmost) end it terminates on the linearly unstable top branch of $(u_0,v_0)$ (Fig.~\ref{fig:6}). This branch also represents a Hopf onset, but there is no bifurcation analogous to that which appears in~(\ref{eq:PDE}). Stable solutions along $TW^-$ branch correspond mostly to $c<0$ (right- and left-moving propagating waves). The sub-critical nature of the $TW^-$ branch agrees with the direct numerical integration of Eq.~(\ref{eq:PDE}): for $Da>Da_W^-$ the emerging states are of large amplitude while for $Da<Da_W^-$ the uniform state is indeed stable to small enough perturbations (Fig.~\ref{fig:3}). The coexistence of stable $TW^\pm$ for $Da^-_W<Da<Da^+_W$ explains the transition from down- to up-stream propagating waves, see Fig.~\ref{fig:3}.  

Similarly to RD systems~\cite{YTDG:08,brena2014subcritical,yochelis2015origin}, and as also indicated by direct numerical integration, the sub-critical regime $Da<Da^-_W$ also supports a multiplicity of secondary solutions with $\lambda \neq \lambda_W^\pm$. While we portray only two of such secondary branches [Fig.~\ref{fig:7}(a)], there are infinitely many solutions , each of which corresponds to a different period. An efficient way of tracking secondary periodic solutions is to compute a nonlinear dispersion relation~\cite{BoEn:03}, i.e., a locus of nonlinear solutions in the ($c,\lambda$) plane at fixed $Da$ value. This is done in the subcritical regime of $TW^-$, since secondary wavenumbers that bifurcate from $Da\gtrsim Da_W^-$ inherit the subcriticality of the primary periodic states~\cite{YTDG:08}. The resulting dispersion relation admits two branches of periodic orbits with positive and negative velocities [see Fig.~\ref{fig:7}(b)]. Notably, the rightmost fold corresponds to downstream propagating $TW$ with periodicity slightly larger than $TW^+$, see point (B) in the top inset in Fig.~\ref{fig:7}(a). Continuation indeed shows that these secondary periodic orbits emerge from $Da>Da_W^-$ and terminate super-critically as small amplitude states at $Da<Da_W^+$. As a comparison, we have followed another periodic orbit with a moderately larger period (marked as (A), as also shown in Fig.~\ref{fig:7}). 

The computation of these periodic orbits remains incomplete,  as it traditionally requires knowledge of pattern selection on large domains, where $L$ is much larger than the critical periodic solutions.  Analysis of temporal stability to long wavelength perturbations ($L=n\lambda,n>1$ so that $L>10$ at least) shows that the stability region shrinks due to bifurcation points accumulating at $Da$ values that are different than those obtained for $L=\lambda$. The small slope of branch (B) in vicinity of $c=0$ agrees with the competition between the upstream $TW^-$ and downstream $TW^+$ families and their slow propagation speed at $Da$ values that are close to $Da^+_W$ [see Fig.~\ref{fig:3}(a)]. In addition, it explains the emergence of upstream moving $TW$, with a period larger than $\lambda_W^+$, that is associated with the coexisting secondary $TW$ solutions, i.e., branch (A) in Fig.~\ref{fig:7}.

\begin{figure*}[tp]
	\includegraphics[width=0.425\textwidth]{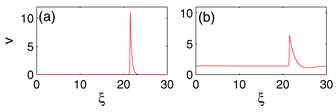}
	\includegraphics[width=0.4\textwidth]{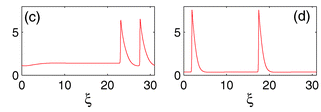}
	\caption{Typical single pulse solutions (a,b) that exist over the entire sub-critical region, double pulse solutions (c) that exist close to $Da^-_W$, and pulse trains (d) that exist at larger distances from $Da^-_W$. {The propagation direction of all solutions is from right to left.} Reprinted from~\cite{yochelis2009towards}.} \label{fig:8}
\end{figure*}

\subsection{Pulses: Waves with infinitely large period}

In addition to TW solutions, Eq.~(\ref{eq:PDE}) admits propagating pulses (solitary waves), which in the context of Eq.~(\ref{eq:ODE}), correspond to homoclinic orbits in space. Figure~\ref{fig:7}(b) shows that both branches in the nonlinear dispersion relation extend to large periods. Solutions that belong to a branch with $c<0$, approach a Shil'nikov- type homoclinic orbit~\cite{GuHo:83}, with monotonic excitation at the front and an oscillatory decay at the rear, see the profile ($\blacklozenge$) in Fig.~\ref{fig:7}(b). Since the pulse shape does not change with an increased period , we regard this solution as a homoclinic orbit, i.e., $\lambda \to \infty$. Indeed linearization about the uniform steady state yields a pair of complex eigenvalues ($\mu_\pm$, with $Re(\mu_\pm)<0$) as well as one real ($\mu_r>0$) where $|Re(\mu_\pm)|<\mu_r$, to a property of Shil'nikov homoclinic orbit. Continuation in $(Da,c)$ with a large fixed period ($\lambda \gg \lambda^-_W$) yields a branch of stable single pulse states ($c<0$) extending over a large interval ($Da<Da^-_W$), while the amplitude of the pulse decreases as $Da$ approaches $Da=Da_W^-$ [profiles (a) and (b) in Fig.~\ref{fig:8}]; this branch is not shown in Fig.~\ref{fig:7}.

The second class of large period solutions, see branch with $c>0$ in the nonlinear dispersion relation, exhibits additions of spatial peaks to the profile as the period increases and therefore is homoclinic to a limit cycle, see profile ($\blacksquare$) in~Fig.~\ref{fig:7}(b). Continuation of this solution below $Da=Da_W^-$, results in bounded states of two peaks with $c<0$, as shown in Fig.~\ref{fig:8}(c). As $Da$ is decreased, the distance between the two pulses increases, as depicted by profile (d).

\begin{figure}[htb]
	\includegraphics[width=0.48\textwidth]{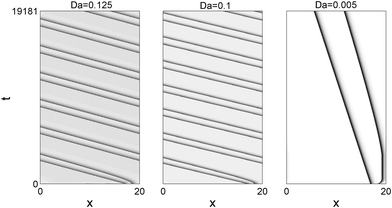}
	\caption{Space-time plots showing the emergences of single and grouped pulses in the excitable regime ($Da<Da_W^-$), where dark color indicates larger $v$ values; Eq.~(\ref{eq:PDE}) was integrated with periodic boundary conditions. The initial perturbation near the outlet includes two confined but well separated large amplitude excitations superimposed on a background of the uniform state. Reprinted from~\cite{yochelis2009towards}.} \label{fig:9}
\end{figure}

In the context of~\eqref{eq:PDE}, solitary waves, which include the aforementioned double peaks, are triggered for $Da<Da_W^-$ via a localized finite amplitude perturbationas demonstrated in Fig.~\ref{fig:9}. For these solutions, direct numerical integrations show that the inter-spacing between pulses increases as $Da$ is decreased. Stability of double pulse states should not come as a surprise, since non-monotonic dispersion relations [see Fig.~\ref{fig:7}(b)] often admit such a property~\cite{EMRS:90}. Under Danckwerts BC, stationary long wavelength bounded states still persist. However, the number of peaks within the bounded state depends on the number of initial perturbations.

\begin{figure}[tp]
	\includegraphics[width=0.35\textwidth]{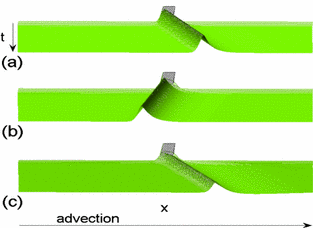}
	\caption{Space-time plots at points that are indicated in Fig.~\ref{fig:1}, showing excitable (a,c) and drifting (b) pulses; Eq.~\ref{eq:PDE} was integrated with no-flux boundary conditions. Reprinted from~\cite{yochelis2010drifting}.} \label{fig:10}
\end{figure}

Surprisingly, solitary waves in RDA system can in fact propagate bidirectionally without changing their profile. Therefore, we distinguish between excitable (upstream) and drifting (downstream) propagations, as shown in Fig.~\ref{fig:10}. The drifting pulses are associated with a convective instability by suppression of the excitation at the fast front ($\xi \to -\infty$) and enhancement of weak deviations at the slow front ($\xi \to \infty$). The drifting pulses exist for $B^*<B<B_0$, and have similar profiles along the stable branch as the standard excitable pulses, see Fig.~\ref{fig:1}. This phenomenon is qualitatively different and cannot occur in a typical RD system ($Le=1$) where the pulses always propagate with a fast excitation at the leading front~\cite{kosek1995splitting,hagberg1998propagation,kiss2004electric}. From physicochemical reasoning, the drifting pulses arise at low reaction-rate regimes of the activator, $Da$, and low exothermicity $B$, see Fig.~\ref{fig:1}. Under such conditions the excitation of nearest neighbors is suppressed due to the advective flow and the drifting pulse is no longer excitable since the leading front now develops from the rest state as a small amplitude perturbation.

Drifting pulses appear to inherit the properties of excitable pulses. The latter are important characteristics of the organization and interaction of solitary waves~\cite{elphick1990impulse,or2000stable,roder2007wave}, and are detected here around $B=B_b$, the Belyakov point~\cite{Bel:74,Bel:80}, see Fig.~\ref{fig:5}(c). At this point, and with an appropriate speed, the spatial eigenvalues correspond to one positive real (associated with $\xi \to -\infty$) and a degenerate pair of negative reals (associated with $\xi \to \infty$). Below $B_b$, the degeneracy is removed but the eigenvalues remain negative reals (a saddle) while above $B_b$ they become complex conjugated corresponding to a saddle focus (a Shil'nikov-type). The interchange of eigenvalues implies a transition from a monotonic to an oscillatory dispersion relation and a monotonic (in space) approach of the homoclinic orbit to the fixed point as $\xi \to \pm\infty$, which also implies coexistence of bounded-pulse states for $B>B_b$~\cite{elphick1990impulse,or2000stable,roder2007wave}.

\section{Pinning and stationary nonuniform solutions}

Stationary solutions cannot be associated with any temporal instability such as a dispersion relation. Yet, using~(\ref{eq:ODE}), SP solutions are found to also correspond to a Hopf bifurcation, but with $c=0$. In the context of~(\ref{eq:PDE}), both onsets $Da=Da_H^\pm$ satisfy the condition $Re[\sigma(k^\pm_H)]=Im[\sigma(k^\pm_H)]=0$~\cite{NNRS:00} (see also Fig.~\ref{fig:6super}), where $k=k_H$ is related to the pair of imaginary spatial eigenvalues, $\mu_\pm=\pm ik^+_H$, at $Da=Da_H^+$ and $\mu_\pm=\pm ik^-_H$, at $Da=Da_H^-$.

\begin{figure}[tp]
	\includegraphics[width=0.4\textwidth]{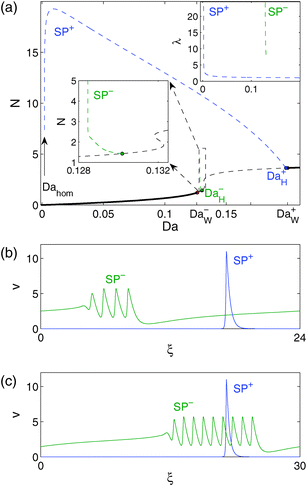}
	\caption{(a) Bifurcation diagram for spatially nonuniform steady states $SP^\pm$ that emerge respectively from $Da_H^\pm$, as a function of $Da$. Eq.~(\ref{eq:ODE}) was integrated on periodic domains with $c=0$. The $SP^+$ solutions approach a homoclinic orbit to a fixed point, $\lambda\to \infty$ as $Da \to Da_{hom}\simeq 0$, while the $SP^-$ solutions approach a homoclinic orbit to a limit cycle. The left inset magnifies the respective region of the $SP^-$ branch. The right inset presents the $SP^\pm$ branches in terms of a spatial period, $\lambda$. (b-c) Profiles of $SP^\pm$ states on large domains, respectively. Reprinted from~\cite{yochelis2009towards}.} \label{fig:11}
\end{figure}

The branch of spatially periodic orbits ($SP^\pm$) bifurcates from $Da=Da_H^\pm \simeq Da_W^\pm$  respectively [see Fig.~\ref{fig:11}] and inherit the respective properties of the $TW^\pm$ branches. These branches are marked by dashed lines, since, in the case of (\ref{eq:PDE}) with periodic BC, these states inherit the instability of the steady state $(u_0,v_0)$. The top inset in Fig.~\ref{fig:11}(a) shows that the period of the $SP^+$ solutions slowly increases as $Da$ is decreased,  and as $Da$ approaches $Da_{hom}\simeq 0$, there is a rapid increase in the period toward a homoclinic orbit, as demonstrated in the respective profiles in Figs.~\ref{fig:11}(b,c). As the SP solutions approach the homoclinic orbit, they pass through the Belyakov bifurcation [Fig.~\ref{fig:5}(c)]. While this allows for multi-pulse stationary states as characteristic solutions of~(\ref{eq:PDE}), evidently they can be stable only on non-periodic domains and near the inlet.

\begin{figure*}[tp]
	\includegraphics[width=0.9\textwidth]{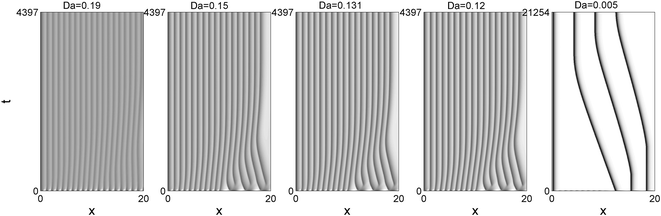}
	\caption{Direct numerical integration of (\ref{eq:PDE}) showing space-time plots for different values of $Da$ and with Danckwerts BC, where dark color indicates larger $v$ field values. As opposed to Fig.~\ref{fig:3}(b), the initial condition is spatially extended and is composed of two periods. Reprinted from~\cite{yochelis2009towards}.} \label{fig:12}
\end{figure*}

Figure~\ref{fig:12} shows, via direct numerical integration, that the wavelengths agree with those obtained by spatial dynamics analysis. Specifically, near $Da_{hom}$ the short wavelength perturbations in the vicinity of the inlet decay to a rest state ($u_0,v_0$), while only long period perturbations, develop to a stationary striped state with a large period.


\section{Summary} 

The objective of the review was to demonstrate the key pattern selection mechanisms that operate on nonlinear patterns in RDA systems with mixed BC on semi-infinite 1D domains. The approach that is most efficient for this purpose is spatial dynamics. This approach, coupled with numerical continuation, deduces the properties of spatially periodic and localized solutions in a co-moving coordinate transformation on periodic domains. Many of the solutions are model independent, since they are organized around global bifurcations, which constrain the system to certain predictable typed of behavior, such as homoclinic orbits. Several of the phenomenon have indeed been observed numerically in other RDA systems, such as FitzHugh--Nagumo~\cite{KM:02,NeSh:03}, Gray--Scott~\cite{SM:00,Sat:03}, and Brusseletor~\cite{KMDB:97} models. {Notably, although RDA systems may  resemble electro-migration of interacting electrically charged species, such as microemulsion and colloids~\cite{bonilla2005non,dahmlow2015electric,strubbe2017charge}, they pose a fundamental difference, since the electrical balance due to Coulombic interactions must take into account also the Poisson equation. The latter framework is thus, belongs to a distinct class of parabolic-elliptic models and while some properties may persist under certain condition, comparison of the two media should be carefully examined~\cite{agladze1992influence,vsevvcikova1999electric,sebestikova2005control,carballido2012effect,dahmlow2015nonlinear}.}

The main results can be summarized as following:
\begin{enumerate}[(i)]
	\item Non-periodic BC, such as Danckwerts type, break the translational symmetry of TW and stabilize SP states, an alternative mechanism  which leads to Turing-type patterns~\cite{yochelis2010turing};
	\item The presence of a sub-critical, finite wavenumber Hopf bifurcation gives rise to upstream propagating TW and solitary waves. The latter are homoclinic orbits in space, which act as generic organizing centers of (nonuniform) spatial solutions
	~\cite{GuHo:83,Kuz:95,SSTC:01}.
\end{enumerate}
We hope that the survey provided here will be useful for exploring in greater detail autocatalytic systems that include a differential flow, high number of variables and diffusing subsets, mass conservation, non-local interactions, higher co-dimension bifurcations and applications thereof~\cite{chomaz1999absolute,KuHo:08,NMTKY:09,sherratt2011pattern,siero2015striped,berenstein2012distinguishing,ghosh2016differential,yochelis2016reaction,holzer2017wavetrain,vidal2017convective,yochelis2015self,gavish2017solvent,ghosh2016differential,siebert2014dynamics,brooks2016mechanism,zmurchok2017application,altimari2012formation,berenstein2012flow,carballido2012effect}. 

\acknowledgments
I am grateful to Sariel Bier for commenting and proof reading the manuscript, and I am also in debt to Moshe Sheintuch, not only for introducing me to this fascinating topic, but also for the productive period that we had jointly worked on it.

\pagebreak


\begin{thebibliography}{108}
	\expandafter\ifx\csname natexlab\endcsname\relax\def\natexlab#1{#1}\fi
	\expandafter\ifx\csname bibnamefont\endcsname\relax
	\def\bibnamefont#1{#1}\fi
	\expandafter\ifx\csname bibfnamefont\endcsname\relax
	\def\bibfnamefont#1{#1}\fi
	\expandafter\ifx\csname citenamefont\endcsname\relax
	\def\citenamefont#1{#1}\fi
	\expandafter\ifx\csname url\endcsname\relax
	\def\url#1{\texttt{#1}}\fi
	\expandafter\ifx\csname urlprefix\endcsname\relax\def\urlprefix{URL }\fi
	\providecommand{\bibinfo}[2]{#2}
	\providecommand{\eprint}[2][]{\url{#2}}
	
	\bibitem[{\citenamefont{Cross and Hohenberg}(1993)}]{CrHo:93}
	\bibinfo{author}{\bibfnamefont{M.~C.} \bibnamefont{Cross}} \bibnamefont{and}
	\bibinfo{author}{\bibfnamefont{P.~C.} \bibnamefont{Hohenberg}},
	\bibinfo{journal}{Reviews of Modern Physics} \textbf{\bibinfo{volume}{65}},
	\bibinfo{pages}{851} (\bibinfo{year}{1993}).
	
	\bibitem[{\citenamefont{Maini et~al.}(1997)\citenamefont{Maini, Painter, and
			Chau}}]{MPC:97}
	\bibinfo{author}{\bibfnamefont{P.}~\bibnamefont{Maini}},
	\bibinfo{author}{\bibfnamefont{K.}~\bibnamefont{Painter}}, \bibnamefont{and}
	\bibinfo{author}{\bibfnamefont{H.~P.} \bibnamefont{Chau}},
	\bibinfo{journal}{Journal of the Chemical Society, Faraday Transactions}
	\textbf{\bibinfo{volume}{93}}, \bibinfo{pages}{3601} (\bibinfo{year}{1997}).
	
	\bibitem[{\citenamefont{Kapral and Showalter}(2012)}]{kapral2012chemical}
	\bibinfo{author}{\bibfnamefont{R.}~\bibnamefont{Kapral}} \bibnamefont{and}
	\bibinfo{author}{\bibfnamefont{K.}~\bibnamefont{Showalter}},
	\emph{\bibinfo{title}{Chemical waves and patterns}},
	vol.~\bibinfo{volume}{10} (\bibinfo{publisher}{Springer Science \& Business
		Media}, \bibinfo{year}{2012}).
	
	\bibitem[{\citenamefont{Turing}(1952)}]{Tur:52}
	\bibinfo{author}{\bibfnamefont{A.~M.} \bibnamefont{Turing}},
	\bibinfo{journal}{Philosophical Transactions of the Royal Society of London
		B: Biological Sciences} \textbf{\bibinfo{volume}{237}}, \bibinfo{pages}{37}
	(\bibinfo{year}{1952}).
	
	\bibitem[{\citenamefont{Keener and Sneyd}(1998)}]{keener1998mathematical}
	\bibinfo{author}{\bibfnamefont{J.}~\bibnamefont{Keener}} \bibnamefont{and}
	\bibinfo{author}{\bibfnamefont{J.}~\bibnamefont{Sneyd}},
	\emph{\bibinfo{title}{Mathematical physiology. interdisciplinary applied
			mathematics, vol. 8}} (\bibinfo{year}{1998}).
	
	\bibitem[{\citenamefont{Murray}(2001)}]{Mu:02}
	\bibinfo{author}{\bibfnamefont{J.~D.} \bibnamefont{Murray}},
	\emph{\bibinfo{title}{Mathematical Biology. II Spatial Models and Biomedical
			Applications $\{$Interdisciplinary Applied Mathematics V. 18$\}$}}
	(\bibinfo{publisher}{Springer-Verlag New York Incorporated},
	\bibinfo{year}{2001}).
	
	\bibitem[{\citenamefont{Pismen}(2006)}]{Pis:06}
	\bibinfo{author}{\bibfnamefont{L.~M.} \bibnamefont{Pismen}},
	\emph{\bibinfo{title}{Patterns and Interfaces in Dissipative Dynamics}}
	(\bibinfo{publisher}{Berlin, Springer}, \bibinfo{year}{2006}).
	
	\bibitem[{\citenamefont{Meron}(2015)}]{Meron2015}
	\bibinfo{author}{\bibfnamefont{E.}~\bibnamefont{Meron}},
	\emph{\bibinfo{title}{Nonlinear Physics of Ecosystems}}
	(\bibinfo{publisher}{Taylor \& Francis Group}, \bibinfo{address}{CRC Press},
	\bibinfo{year}{2015}).
	
	\bibitem[{\citenamefont{Hoyle}(2006)}]{Hoyle:06}
	\bibinfo{author}{\bibfnamefont{R.~B.} \bibnamefont{Hoyle}},
	\emph{\bibinfo{title}{Pattern Formation: An Introduction to Methods}}
	(\bibinfo{publisher}{Cambridge University Press, Cambridge},
	\bibinfo{year}{2006}).
	
	\bibitem[{\citenamefont{Knobloch}(2002)}]{Kn:02}
	\bibinfo{author}{\bibfnamefont{E.}~\bibnamefont{Knobloch}},
	\emph{\bibinfo{title}{Nonlinear Dynamics and Chaos: Where do we go from
			here?}} (\bibinfo{publisher}{CRC Press}, \bibinfo{year}{2002}).
	
	\bibitem[{\citenamefont{Huerre and Monkewitz}(1990)}]{HuMo:90}
	\bibinfo{author}{\bibfnamefont{P.}~\bibnamefont{Huerre}} \bibnamefont{and}
	\bibinfo{author}{\bibfnamefont{P.~A.} \bibnamefont{Monkewitz}},
	\bibinfo{journal}{Annual Review of Fluid Mechanics}
	\textbf{\bibinfo{volume}{22}}, \bibinfo{pages}{473} (\bibinfo{year}{1990}).
	
	\bibitem[{\citenamefont{Chomaz}(1992)}]{Chom:92}
	\bibinfo{author}{\bibfnamefont{J.}~\bibnamefont{Chomaz}},
	\bibinfo{journal}{Physical Review Letters} \textbf{\bibinfo{volume}{69}},
	\bibinfo{pages}{1931} (\bibinfo{year}{1992}).
	
	\bibitem[{\citenamefont{Yakhnin et~al.}(1995)\citenamefont{Yakhnin, Rovinsky,
			and Menzinger}}]{yakhnin1995convective}
	\bibinfo{author}{\bibfnamefont{V.}~\bibnamefont{Yakhnin}},
	\bibinfo{author}{\bibfnamefont{A.}~\bibnamefont{Rovinsky}}, \bibnamefont{and}
	\bibinfo{author}{\bibfnamefont{M.}~\bibnamefont{Menzinger}},
	\bibinfo{journal}{Chemical Engineering Science}
	\textbf{\bibinfo{volume}{50}}, \bibinfo{pages}{2853} (\bibinfo{year}{1995}).
	
	\bibitem[{\citenamefont{Khazan and Pismen}(1995)}]{KhPi:95}
	\bibinfo{author}{\bibfnamefont{Y.}~\bibnamefont{Khazan}} \bibnamefont{and}
	\bibinfo{author}{\bibfnamefont{L.}~\bibnamefont{Pismen}},
	\bibinfo{journal}{Physical Review Letters} \textbf{\bibinfo{volume}{75}},
	\bibinfo{pages}{4318} (\bibinfo{year}{1995}).
	
	\bibitem[{\citenamefont{Kosek et~al.}(1995)\citenamefont{Kosek, Sevcikova, and
			Marek}}]{kosek1995splitting}
	\bibinfo{author}{\bibfnamefont{J.}~\bibnamefont{Kosek}},
	\bibinfo{author}{\bibfnamefont{H.}~\bibnamefont{Sevcikova}},
	\bibnamefont{and} \bibinfo{author}{\bibfnamefont{M.}~\bibnamefont{Marek}},
	\bibinfo{journal}{The Journal of Physical Chemistry}
	\textbf{\bibinfo{volume}{99}}, \bibinfo{pages}{6889} (\bibinfo{year}{1995}).
	
	\bibitem[{\citenamefont{Sheintuch}(1997)}]{Sh:97}
	\bibinfo{author}{\bibfnamefont{M.}~\bibnamefont{Sheintuch}},
	\bibinfo{journal}{Physica D} \textbf{\bibinfo{volume}{102}},
	\bibinfo{pages}{125} (\bibinfo{year}{1997}).
	
	\bibitem[{\citenamefont{Kuznetsov et~al.}(1997)\citenamefont{Kuznetsov,
			Mosekilde, Dewel, and Borckmans}}]{KMDB:97}
	\bibinfo{author}{\bibfnamefont{S.~P.} \bibnamefont{Kuznetsov}},
	\bibinfo{author}{\bibfnamefont{E.}~\bibnamefont{Mosekilde}},
	\bibinfo{author}{\bibfnamefont{G.}~\bibnamefont{Dewel}}, \bibnamefont{and}
	\bibinfo{author}{\bibfnamefont{P.}~\bibnamefont{Borckmans}},
	\bibinfo{journal}{The Journal of chemical physics}
	\textbf{\bibinfo{volume}{106}}, \bibinfo{pages}{7609} (\bibinfo{year}{1997}).
	
	\bibitem[{\citenamefont{Satnoianu et~al.}(1998)\citenamefont{Satnoianu, Merkin,
			and Scott}}]{SMS:98}
	\bibinfo{author}{\bibfnamefont{R.~A.} \bibnamefont{Satnoianu}},
	\bibinfo{author}{\bibfnamefont{J.~H.} \bibnamefont{Merkin}},
	\bibnamefont{and} \bibinfo{author}{\bibfnamefont{S.~K.} \bibnamefont{Scott}},
	\bibinfo{journal}{Physica D} \textbf{\bibinfo{volume}{124}},
	\bibinfo{pages}{345} (\bibinfo{year}{1998}).
	
	\bibitem[{\citenamefont{Andres{\'e}n et~al.}(1999)\citenamefont{Andres{\'e}n,
			Bache, Mosekilde, Dewel, and Borckmanns}}]{ABMDB:99}
	\bibinfo{author}{\bibfnamefont{P.}~\bibnamefont{Andres{\'e}n}},
	\bibinfo{author}{\bibfnamefont{M.}~\bibnamefont{Bache}},
	\bibinfo{author}{\bibfnamefont{E.}~\bibnamefont{Mosekilde}},
	\bibinfo{author}{\bibfnamefont{G.}~\bibnamefont{Dewel}}, \bibnamefont{and}
	\bibinfo{author}{\bibfnamefont{P.}~\bibnamefont{Borckmanns}},
	\bibinfo{journal}{Physical Review E} \textbf{\bibinfo{volume}{60}},
	\bibinfo{pages}{297} (\bibinfo{year}{1999}).
	
	\bibitem[{\citenamefont{Nekhamkina
			et~al.}(2000{\natexlab{a}})\citenamefont{Nekhamkina, Rubinstein, and
			Sheintuch}}]{NRS:00}
	\bibinfo{author}{\bibfnamefont{O.}~\bibnamefont{Nekhamkina}},
	\bibinfo{author}{\bibfnamefont{B.~Y.} \bibnamefont{Rubinstein}},
	\bibnamefont{and}
	\bibinfo{author}{\bibfnamefont{M.}~\bibnamefont{Sheintuch}},
	\bibinfo{journal}{AIChE Journal} \textbf{\bibinfo{volume}{46}},
	\bibinfo{pages}{1632} (\bibinfo{year}{2000}{\natexlab{a}}).
	
	\bibitem[{\citenamefont{Satnoianu and Menzinger}(2000)}]{SM:00}
	\bibinfo{author}{\bibfnamefont{R.~A.} \bibnamefont{Satnoianu}}
	\bibnamefont{and}
	\bibinfo{author}{\bibfnamefont{M.}~\bibnamefont{Menzinger}},
	\bibinfo{journal}{Physical Review E} \textbf{\bibinfo{volume}{62}},
	\bibinfo{pages}{113} (\bibinfo{year}{2000}).
	
	\bibitem[{\citenamefont{Bamforth et~al.}(2000)\citenamefont{Bamforth,
			Kalliadasis, Merkin, and Scott}}]{BKMS:00}
	\bibinfo{author}{\bibfnamefont{J.~R.} \bibnamefont{Bamforth}},
	\bibinfo{author}{\bibfnamefont{S.}~\bibnamefont{Kalliadasis}},
	\bibinfo{author}{\bibfnamefont{J.~H.} \bibnamefont{Merkin}},
	\bibnamefont{and} \bibinfo{author}{\bibfnamefont{S.~K.} \bibnamefont{Scott}},
	\bibinfo{journal}{Physical Chemistry Chemical Physics}
	\textbf{\bibinfo{volume}{2}}, \bibinfo{pages}{4013} (\bibinfo{year}{2000}).
	
	\bibitem[{\citenamefont{Satnoianu et~al.}(2000)\citenamefont{Satnoianu,
			Menzinger, and Maini}}]{SMM:00}
	\bibinfo{author}{\bibfnamefont{R.~A.} \bibnamefont{Satnoianu}},
	\bibinfo{author}{\bibfnamefont{M.}~\bibnamefont{Menzinger}},
	\bibnamefont{and} \bibinfo{author}{\bibfnamefont{P.~K.} \bibnamefont{Maini}},
	\bibinfo{journal}{Journal of Mathematical Biology}
	\textbf{\bibinfo{volume}{41}}, \bibinfo{pages}{493} (\bibinfo{year}{2000}).
	
	\bibitem[{\citenamefont{Satnoianu et~al.}(2001)\citenamefont{Satnoianu, Maini,
			and Menzinger}}]{SMM:01}
	\bibinfo{author}{\bibfnamefont{R.~A.} \bibnamefont{Satnoianu}},
	\bibinfo{author}{\bibfnamefont{P.~K.} \bibnamefont{Maini}}, \bibnamefont{and}
	\bibinfo{author}{\bibfnamefont{M.}~\bibnamefont{Menzinger}},
	\bibinfo{journal}{Physica D} \textbf{\bibinfo{volume}{160}},
	\bibinfo{pages}{79} (\bibinfo{year}{2001}).
	
	\bibitem[{\citenamefont{Bamforth et~al.}(2001)\citenamefont{Bamforth, Merkin,
			Scott, Toth, and Gaspar}}]{Bam:01}
	\bibinfo{author}{\bibfnamefont{J.}~\bibnamefont{Bamforth}},
	\bibinfo{author}{\bibfnamefont{J.}~\bibnamefont{Merkin}},
	\bibinfo{author}{\bibfnamefont{S.}~\bibnamefont{Scott}},
	\bibinfo{author}{\bibfnamefont{R.}~\bibnamefont{Toth}}, \bibnamefont{and}
	\bibinfo{author}{\bibfnamefont{V.}~\bibnamefont{Gaspar}},
	\bibinfo{journal}{Physical Chemistry Chemical Physics}
	\textbf{\bibinfo{volume}{3}}, \bibinfo{pages}{1435} (\bibinfo{year}{2001}).
	
	\bibitem[{\citenamefont{Nekhamkina and Sheintuch}(2002)}]{NeSh:02}
	\bibinfo{author}{\bibfnamefont{O.}~\bibnamefont{Nekhamkina}} \bibnamefont{and}
	\bibinfo{author}{\bibfnamefont{M.}~\bibnamefont{Sheintuch}},
	\bibinfo{journal}{Physical Review E} \textbf{\bibinfo{volume}{66}},
	\bibinfo{pages}{016204} (\bibinfo{year}{2002}).
	
	\bibitem[{\citenamefont{K{\ae}rn and Menzinger}(2002)}]{KM:02}
	\bibinfo{author}{\bibfnamefont{M.}~\bibnamefont{K{\ae}rn}} \bibnamefont{and}
	\bibinfo{author}{\bibfnamefont{M.}~\bibnamefont{Menzinger}},
	\bibinfo{journal}{Physical Review E} \textbf{\bibinfo{volume}{65}},
	\bibinfo{pages}{046202} (\bibinfo{year}{2002}).
	
	\bibitem[{\citenamefont{Sheintuch and Nekhamkina}(2003)}]{ShNe:03}
	\bibinfo{author}{\bibfnamefont{M.}~\bibnamefont{Sheintuch}} \bibnamefont{and}
	\bibinfo{author}{\bibfnamefont{O.}~\bibnamefont{Nekhamkina}},
	\bibinfo{journal}{AIChE Journal} \textbf{\bibinfo{volume}{49}},
	\bibinfo{pages}{1241} (\bibinfo{year}{2003}).
	
	\bibitem[{\citenamefont{Satnoianu}(2003)}]{Sat:03}
	\bibinfo{author}{\bibfnamefont{R.~A.} \bibnamefont{Satnoianu}},
	\bibinfo{journal}{Physical Review E} \textbf{\bibinfo{volume}{68}},
	\bibinfo{pages}{032101} (\bibinfo{year}{2003}).
	
	\bibitem[{\citenamefont{Nekhamkina and Sheintuch}(2003)}]{NeSh:03}
	\bibinfo{author}{\bibfnamefont{O.}~\bibnamefont{Nekhamkina}} \bibnamefont{and}
	\bibinfo{author}{\bibfnamefont{M.}~\bibnamefont{Sheintuch}},
	\bibinfo{journal}{Physical Review E} \textbf{\bibinfo{volume}{68}},
	\bibinfo{pages}{036207} (\bibinfo{year}{2003}).
	
	\bibitem[{\citenamefont{McGraw and Menzinger}(2005)}]{MM:05}
	\bibinfo{author}{\bibfnamefont{P.~N.} \bibnamefont{McGraw}} \bibnamefont{and}
	\bibinfo{author}{\bibfnamefont{M.}~\bibnamefont{Menzinger}},
	\bibinfo{journal}{Physical Review E} \textbf{\bibinfo{volume}{72}},
	\bibinfo{pages}{015101} (\bibinfo{year}{2005}).
	
	\bibitem[{\citenamefont{M{\'\i}guez et~al.}(2006)\citenamefont{M{\'\i}guez,
			Satnoianu, and Mu{\~n}uzuri}}]{MSM:06}
	\bibinfo{author}{\bibfnamefont{D.~G.} \bibnamefont{M{\'\i}guez}},
	\bibinfo{author}{\bibfnamefont{R.~A.} \bibnamefont{Satnoianu}},
	\bibnamefont{and} \bibinfo{author}{\bibfnamefont{A.~P.}
		\bibnamefont{Mu{\~n}uzuri}}, \bibinfo{journal}{Physical Review E}
	\textbf{\bibinfo{volume}{73}}, \bibinfo{pages}{025201}
	(\bibinfo{year}{2006}).
	
	\bibitem[{\citenamefont{Zhang et~al.}(2006)\citenamefont{Zhang, Mangold, and
			Kienle}}]{ZMK:06}
	\bibinfo{author}{\bibfnamefont{F.}~\bibnamefont{Zhang}},
	\bibinfo{author}{\bibfnamefont{M.}~\bibnamefont{Mangold}}, \bibnamefont{and}
	\bibinfo{author}{\bibfnamefont{A.}~\bibnamefont{Kienle}},
	\bibinfo{journal}{Chemical Engineering Science}
	\textbf{\bibinfo{volume}{61}}, \bibinfo{pages}{7161} (\bibinfo{year}{2006}).
	
	\bibitem[{\citenamefont{Flach et~al.}(2007)\citenamefont{Flach, Schnell, and
			Norbury}}]{FSN:07}
	\bibinfo{author}{\bibfnamefont{E.}~\bibnamefont{Flach}},
	\bibinfo{author}{\bibfnamefont{S.}~\bibnamefont{Schnell}}, \bibnamefont{and}
	\bibinfo{author}{\bibfnamefont{J.}~\bibnamefont{Norbury}},
	\bibinfo{journal}{Physical Review E} \textbf{\bibinfo{volume}{76}},
	\bibinfo{pages}{036216} (\bibinfo{year}{2007}).
	
	\bibitem[{\citenamefont{Yamada et~al.}(2007)\citenamefont{Yamada, Nakagaki,
			Baker, and Maini}}]{YNBM:07}
	\bibinfo{author}{\bibfnamefont{H.}~\bibnamefont{Yamada}},
	\bibinfo{author}{\bibfnamefont{T.}~\bibnamefont{Nakagaki}},
	\bibinfo{author}{\bibfnamefont{R.~E.} \bibnamefont{Baker}}, \bibnamefont{and}
	\bibinfo{author}{\bibfnamefont{P.~K.} \bibnamefont{Maini}},
	\bibinfo{journal}{Journal of Mathematical Biology}
	\textbf{\bibinfo{volume}{54}}, \bibinfo{pages}{745} (\bibinfo{year}{2007}).
	
	\bibitem[{\citenamefont{Vasquez et~al.}(2008)\citenamefont{Vasquez, Meyer, and
			Suedhoff}}]{VMS:08}
	\bibinfo{author}{\bibfnamefont{D.~A.} \bibnamefont{Vasquez}},
	\bibinfo{author}{\bibfnamefont{J.}~\bibnamefont{Meyer}}, \bibnamefont{and}
	\bibinfo{author}{\bibfnamefont{H.}~\bibnamefont{Suedhoff}},
	\bibinfo{journal}{Physical Review E} \textbf{\bibinfo{volume}{78}},
	\bibinfo{pages}{036109} (\bibinfo{year}{2008}).
	
	\bibitem[{\citenamefont{Sheintuch and
			Shvartsman}(1996)}]{sheintuch1996spatiotemporal}
	\bibinfo{author}{\bibfnamefont{M.}~\bibnamefont{Sheintuch}} \bibnamefont{and}
	\bibinfo{author}{\bibfnamefont{S.}~\bibnamefont{Shvartsman}},
	\bibinfo{journal}{AIChE journal} \textbf{\bibinfo{volume}{42}},
	\bibinfo{pages}{1041} (\bibinfo{year}{1996}).
	
	\bibitem[{\citenamefont{K{\ae}rn et~al.}(2002)\citenamefont{K{\ae}rn,
			Menzinger, Satnoianu, and Hunding}}]{KMSH:02}
	\bibinfo{author}{\bibfnamefont{M.}~\bibnamefont{K{\ae}rn}},
	\bibinfo{author}{\bibfnamefont{M.}~\bibnamefont{Menzinger}},
	\bibinfo{author}{\bibfnamefont{R.}~\bibnamefont{Satnoianu}},
	\bibnamefont{and} \bibinfo{author}{\bibfnamefont{A.}~\bibnamefont{Hunding}},
	\bibinfo{journal}{Faraday Discussions} \textbf{\bibinfo{volume}{120}},
	\bibinfo{pages}{295} (\bibinfo{year}{2002}).
	
	\bibitem[{\citenamefont{Borgogno et~al.}(2009)\citenamefont{Borgogno,
			D'Odorico, Laio, and Ridolfi}}]{BDLR:09}
	\bibinfo{author}{\bibfnamefont{F.}~\bibnamefont{Borgogno}},
	\bibinfo{author}{\bibfnamefont{P.}~\bibnamefont{D'Odorico}},
	\bibinfo{author}{\bibfnamefont{F.}~\bibnamefont{Laio}}, \bibnamefont{and}
	\bibinfo{author}{\bibfnamefont{L.}~\bibnamefont{Ridolfi}},
	\bibinfo{journal}{Reviews of Geophysics} \textbf{\bibinfo{volume}{47}}
	(\bibinfo{year}{2009}).
	
	\bibitem[{\citenamefont{Rovinsky and Menzinger}(1993)}]{RoMe:93}
	\bibinfo{author}{\bibfnamefont{A.~B.} \bibnamefont{Rovinsky}} \bibnamefont{and}
	\bibinfo{author}{\bibfnamefont{M.}~\bibnamefont{Menzinger}},
	\bibinfo{journal}{Physical Review Letters} \textbf{\bibinfo{volume}{70}},
	\bibinfo{pages}{778} (\bibinfo{year}{1993}).
	
	\bibitem[{\citenamefont{Klausmeier}(1999)}]{Klaus:99}
	\bibinfo{author}{\bibfnamefont{C.~A.} \bibnamefont{Klausmeier}},
	\bibinfo{journal}{Science} \textbf{\bibinfo{volume}{284}},
	\bibinfo{pages}{1826} (\bibinfo{year}{1999}).
	
	\bibitem[{\citenamefont{K{\ae}rn and Menzinger}(1999)}]{KM:99}
	\bibinfo{author}{\bibfnamefont{M.}~\bibnamefont{K{\ae}rn}} \bibnamefont{and}
	\bibinfo{author}{\bibfnamefont{M.}~\bibnamefont{Menzinger}},
	\bibinfo{journal}{Physical Review E} \textbf{\bibinfo{volume}{60}},
	\bibinfo{pages}{R3471} (\bibinfo{year}{1999}).
	
	\bibitem[{\citenamefont{Nekhamkina
			et~al.}(2000{\natexlab{b}})\citenamefont{Nekhamkina, Nepomnyashchy,
			Rubinstein, and Sheintuch}}]{NNRS:00}
	\bibinfo{author}{\bibfnamefont{O.~A.} \bibnamefont{Nekhamkina}},
	\bibinfo{author}{\bibfnamefont{A.~A.} \bibnamefont{Nepomnyashchy}},
	\bibinfo{author}{\bibfnamefont{B.~Y.} \bibnamefont{Rubinstein}},
	\bibnamefont{and}
	\bibinfo{author}{\bibfnamefont{M.}~\bibnamefont{Sheintuch}},
	\bibinfo{journal}{Physical Review E} \textbf{\bibinfo{volume}{61}},
	\bibinfo{pages}{2436} (\bibinfo{year}{2000}{\natexlab{b}}).
	
	\bibitem[{\citenamefont{Couairon and Chomaz}(1999)}]{couairon1999primary}
	\bibinfo{author}{\bibfnamefont{A.}~\bibnamefont{Couairon}} \bibnamefont{and}
	\bibinfo{author}{\bibfnamefont{J.-M.} \bibnamefont{Chomaz}},
	\bibinfo{journal}{Physica D} \textbf{\bibinfo{volume}{132}},
	\bibinfo{pages}{428} (\bibinfo{year}{1999}).
	
	\bibitem[{\citenamefont{Deissler}(1985)}]{deissler1985noise}
	\bibinfo{author}{\bibfnamefont{R.~J.} \bibnamefont{Deissler}},
	\bibinfo{journal}{Journal of Statistical Physics}
	\textbf{\bibinfo{volume}{40}}, \bibinfo{pages}{371} (\bibinfo{year}{1985}).
	
	\bibitem[{\citenamefont{M{\"u}ller and Tveitereid}(1995)}]{MuTv:95}
	\bibinfo{author}{\bibfnamefont{H.~W.} \bibnamefont{M{\"u}ller}}
	\bibnamefont{and}
	\bibinfo{author}{\bibfnamefont{M.}~\bibnamefont{Tveitereid}},
	\bibinfo{journal}{Physical Review Letters} \textbf{\bibinfo{volume}{74}},
	\bibinfo{pages}{1582} (\bibinfo{year}{1995}).
	
	\bibitem[{\citenamefont{Couairon and Chomaz}(1997)}]{CoCh:97}
	\bibinfo{author}{\bibfnamefont{A.}~\bibnamefont{Couairon}} \bibnamefont{and}
	\bibinfo{author}{\bibfnamefont{J.}~\bibnamefont{Chomaz}},
	\bibinfo{journal}{Physical Review Letters} \textbf{\bibinfo{volume}{79}},
	\bibinfo{pages}{2666} (\bibinfo{year}{1997}).
	
	\bibitem[{\citenamefont{Tobias et~al.}(1998)\citenamefont{Tobias, Proctor, and
			Knobloch}}]{TPK:98}
	\bibinfo{author}{\bibfnamefont{S.}~\bibnamefont{Tobias}},
	\bibinfo{author}{\bibfnamefont{M.}~\bibnamefont{Proctor}}, \bibnamefont{and}
	\bibinfo{author}{\bibfnamefont{E.}~\bibnamefont{Knobloch}},
	\bibinfo{journal}{Physica D} \textbf{\bibinfo{volume}{113}},
	\bibinfo{pages}{43} (\bibinfo{year}{1998}).
	
	\bibitem[{\citenamefont{Yochelis and
			Sheintuch}(2009{\natexlab{a}})}]{yochelis2009principal}
	\bibinfo{author}{\bibfnamefont{A.}~\bibnamefont{Yochelis}} \bibnamefont{and}
	\bibinfo{author}{\bibfnamefont{M.}~\bibnamefont{Sheintuch}},
	\bibinfo{journal}{Physical Review E} \textbf{\bibinfo{volume}{80}},
	\bibinfo{pages}{056201} (\bibinfo{year}{2009}{\natexlab{a}}).
	
	\bibitem[{\citenamefont{Yochelis and
			Sheintuch}(2009{\natexlab{b}})}]{yochelis2009towards}
	\bibinfo{author}{\bibfnamefont{A.}~\bibnamefont{Yochelis}} \bibnamefont{and}
	\bibinfo{author}{\bibfnamefont{M.}~\bibnamefont{Sheintuch}},
	\bibinfo{journal}{Physical Chemistry Chemical Physics}
	\textbf{\bibinfo{volume}{11}}, \bibinfo{pages}{9210}
	(\bibinfo{year}{2009}{\natexlab{b}}).
	
	\bibitem[{\citenamefont{Yochelis and
			Sheintuch}(2010{\natexlab{a}})}]{yochelis2010drifting}
	\bibinfo{author}{\bibfnamefont{A.}~\bibnamefont{Yochelis}} \bibnamefont{and}
	\bibinfo{author}{\bibfnamefont{M.}~\bibnamefont{Sheintuch}},
	\bibinfo{journal}{Physical Review E} \textbf{\bibinfo{volume}{81}},
	\bibinfo{pages}{025203} (\bibinfo{year}{2010}{\natexlab{a}}).
	
	\bibitem[{\citenamefont{Yakhnin
			et~al.}(1994{\natexlab{a}})\citenamefont{Yakhnin, Rovinsky, and
			Menzinger}}]{yakhnin1994differential}
	\bibinfo{author}{\bibfnamefont{V.}~\bibnamefont{Yakhnin}},
	\bibinfo{author}{\bibfnamefont{A.}~\bibnamefont{Rovinsky}}, \bibnamefont{and}
	\bibinfo{author}{\bibfnamefont{M.}~\bibnamefont{Menzinger}},
	\bibinfo{journal}{Chemical Engineering Science}
	\textbf{\bibinfo{volume}{49}}, \bibinfo{pages}{3257}
	(\bibinfo{year}{1994}{\natexlab{a}}).
	
	\bibitem[{\citenamefont{Yakhnin
			et~al.}(1994{\natexlab{b}})\citenamefont{Yakhnin, Rovinsky, and
			Menzinger}}]{yakhnin1994differential_b}
	\bibinfo{author}{\bibfnamefont{V.~Z.} \bibnamefont{Yakhnin}},
	\bibinfo{author}{\bibfnamefont{A.~B.} \bibnamefont{Rovinsky}},
	\bibnamefont{and}
	\bibinfo{author}{\bibfnamefont{M.}~\bibnamefont{Menzinger}},
	\bibinfo{journal}{Journal of Physical Chemistry}
	\textbf{\bibinfo{volume}{98}}, \bibinfo{pages}{2116}
	(\bibinfo{year}{1994}{\natexlab{b}}).
	
	\bibitem[{\citenamefont{Sheintuch and Nekhamkina}(1999)}]{ShNe:99}
	\bibinfo{author}{\bibfnamefont{M.}~\bibnamefont{Sheintuch}} \bibnamefont{and}
	\bibinfo{author}{\bibfnamefont{O.}~\bibnamefont{Nekhamkina}},
	\bibinfo{journal}{AIChE Journal} \textbf{\bibinfo{volume}{45}},
	\bibinfo{pages}{398} (\bibinfo{year}{1999}).
	
	\bibitem[{\citenamefont{Uppal et~al.}(1974)\citenamefont{Uppal, Ray, and
			Poore}}]{URP:74}
	\bibinfo{author}{\bibfnamefont{A.}~\bibnamefont{Uppal}},
	\bibinfo{author}{\bibfnamefont{W.}~\bibnamefont{Ray}}, \bibnamefont{and}
	\bibinfo{author}{\bibfnamefont{A.}~\bibnamefont{Poore}},
	\bibinfo{journal}{Chemical Engineering Science}
	\textbf{\bibinfo{volume}{29}}, \bibinfo{pages}{967} (\bibinfo{year}{1974}).
	
	\bibitem[{\citenamefont{Froment et~al.}(2011)\citenamefont{Froment, Bischoff,
			and De~Wilde}}]{froment2011chemical}
	\bibinfo{author}{\bibfnamefont{G.~F.} \bibnamefont{Froment}},
	\bibinfo{author}{\bibfnamefont{K.~B.} \bibnamefont{Bischoff}},
	\bibnamefont{and} \bibinfo{author}{\bibfnamefont{J.}~\bibnamefont{De~Wilde}},
	\emph{\bibinfo{title}{Chemical Reactor-Analysis and Design}}
	(\bibinfo{year}{2011}).
	
	\bibitem[{\citenamefont{Sheintuch and Nekhamkina}(2005)}]{ShNe:05}
	\bibinfo{author}{\bibfnamefont{M.}~\bibnamefont{Sheintuch}} \bibnamefont{and}
	\bibinfo{author}{\bibfnamefont{O.}~\bibnamefont{Nekhamkina}},
	\bibinfo{journal}{AIChE Journal} \textbf{\bibinfo{volume}{51}},
	\bibinfo{pages}{224} (\bibinfo{year}{2005}).
	
	\bibitem[{\citenamefont{Yochelis
			et~al.}(2008{\natexlab{a}})\citenamefont{Yochelis, Knobloch, Xie, Qu, and
			Garfinkel}}]{Yo:08}
	\bibinfo{author}{\bibfnamefont{A.}~\bibnamefont{Yochelis}},
	\bibinfo{author}{\bibfnamefont{E.}~\bibnamefont{Knobloch}},
	\bibinfo{author}{\bibfnamefont{Y.}~\bibnamefont{Xie}},
	\bibinfo{author}{\bibfnamefont{Z.}~\bibnamefont{Qu}}, \bibnamefont{and}
	\bibinfo{author}{\bibfnamefont{A.}~\bibnamefont{Garfinkel}},
	\bibinfo{journal}{EPL (Europhysics Letters)} \textbf{\bibinfo{volume}{83}},
	\bibinfo{pages}{64005} (\bibinfo{year}{2008}{\natexlab{a}}).
	
	\bibitem[{\citenamefont{Anma et~al.}(2012)\citenamefont{Anma, Sakamoto, and
			Yoneda}}]{anma2012unstable}
	\bibinfo{author}{\bibfnamefont{A.}~\bibnamefont{Anma}},
	\bibinfo{author}{\bibfnamefont{K.}~\bibnamefont{Sakamoto}}, \bibnamefont{and}
	\bibinfo{author}{\bibfnamefont{T.}~\bibnamefont{Yoneda}},
	\bibinfo{journal}{Kodai Mathematical Journal} \textbf{\bibinfo{volume}{35}},
	\bibinfo{pages}{215} (\bibinfo{year}{2012}).
	
	\bibitem[{\citenamefont{Hata et~al.}(2014)\citenamefont{Hata, Nakao, and
			Mikhailov}}]{hata2014sufficient}
	\bibinfo{author}{\bibfnamefont{S.}~\bibnamefont{Hata}},
	\bibinfo{author}{\bibfnamefont{H.}~\bibnamefont{Nakao}}, \bibnamefont{and}
	\bibinfo{author}{\bibfnamefont{A.~S.} \bibnamefont{Mikhailov}},
	\bibinfo{journal}{Progress of Theoretical and Experimental Physics}
	\textbf{\bibinfo{volume}{2014}}, \bibinfo{pages}{1} (\bibinfo{year}{2014}).
	
	\bibitem[{\citenamefont{Knobloch}(1986)}]{knobloch1986oscillatory}
	\bibinfo{author}{\bibfnamefont{E.}~\bibnamefont{Knobloch}},
	\bibinfo{journal}{Physical Review A} \textbf{\bibinfo{volume}{34}},
	\bibinfo{pages}{1538} (\bibinfo{year}{1986}).
	
	\bibitem[{\citenamefont{Burke and Knobloch}(2007)}]{burke2007homoclinic}
	\bibinfo{author}{\bibfnamefont{J.}~\bibnamefont{Burke}} \bibnamefont{and}
	\bibinfo{author}{\bibfnamefont{E.}~\bibnamefont{Knobloch}},
	\bibinfo{journal}{Chaos: An Interdisciplinary Journal of Nonlinear Science}
	\textbf{\bibinfo{volume}{17}}, \bibinfo{pages}{037102}
	(\bibinfo{year}{2007}).
	
	\bibitem[{\citenamefont{Yochelis
			et~al.}(2008{\natexlab{b}})\citenamefont{Yochelis, Tintut, Demer, and
			Garfinkel}}]{YTDG:08}
	\bibinfo{author}{\bibfnamefont{A.}~\bibnamefont{Yochelis}},
	\bibinfo{author}{\bibfnamefont{Y.}~\bibnamefont{Tintut}},
	\bibinfo{author}{\bibfnamefont{L.}~\bibnamefont{Demer}}, \bibnamefont{and}
	\bibinfo{author}{\bibfnamefont{A.}~\bibnamefont{Garfinkel}},
	\bibinfo{journal}{New Journal of Physics} \textbf{\bibinfo{volume}{10}},
	\bibinfo{pages}{055002} (\bibinfo{year}{2008}{\natexlab{b}}).
	
	\bibitem[{\citenamefont{Dawes}(2008)}]{dawes2008localized}
	\bibinfo{author}{\bibfnamefont{J.~H.} \bibnamefont{Dawes}},
	\bibinfo{journal}{SIAM Journal on Applied Dynamical Systems}
	\textbf{\bibinfo{volume}{7}}, \bibinfo{pages}{186} (\bibinfo{year}{2008}).
	
	\bibitem[{\citenamefont{Burke et~al.}(2008)\citenamefont{Burke, Yochelis, and
			Knobloch}}]{BYK:08}
	\bibinfo{author}{\bibfnamefont{J.}~\bibnamefont{Burke}},
	\bibinfo{author}{\bibfnamefont{A.}~\bibnamefont{Yochelis}}, \bibnamefont{and}
	\bibinfo{author}{\bibfnamefont{E.}~\bibnamefont{Knobloch}},
	\bibinfo{journal}{SIAM Journal on Applied Dynamical Systems}
	\textbf{\bibinfo{volume}{7}}, \bibinfo{pages}{651} (\bibinfo{year}{2008}).
	
	\bibitem[{\citenamefont{Kozyreff et~al.}(2009)\citenamefont{Kozyreff, Assemat,
			and Chapman}}]{kozyreff2009influence}
	\bibinfo{author}{\bibfnamefont{G.}~\bibnamefont{Kozyreff}},
	\bibinfo{author}{\bibfnamefont{P.}~\bibnamefont{Assemat}}, \bibnamefont{and}
	\bibinfo{author}{\bibfnamefont{S.~J.} \bibnamefont{Chapman}},
	\bibinfo{journal}{Physical Review Letters} \textbf{\bibinfo{volume}{103}},
	\bibinfo{pages}{164501} (\bibinfo{year}{2009}).
	
	\bibitem[{\citenamefont{Yochelis
			et~al.}(2015{\natexlab{a}})\citenamefont{Yochelis, Knobloch, and
			K{\"o}pf}}]{yochelis2015origin}
	\bibinfo{author}{\bibfnamefont{A.}~\bibnamefont{Yochelis}},
	\bibinfo{author}{\bibfnamefont{E.}~\bibnamefont{Knobloch}}, \bibnamefont{and}
	\bibinfo{author}{\bibfnamefont{M.~H.} \bibnamefont{K{\"o}pf}},
	\bibinfo{journal}{Physical Review E} \textbf{\bibinfo{volume}{91}},
	\bibinfo{pages}{032924} (\bibinfo{year}{2015}{\natexlab{a}}).
	
	\bibitem[{\citenamefont{Thiele et~al.}(2013)\citenamefont{Thiele, Archer,
			Robbins, Gomez, and Knobloch}}]{thiele2013localized}
	\bibinfo{author}{\bibfnamefont{U.}~\bibnamefont{Thiele}},
	\bibinfo{author}{\bibfnamefont{A.~J.} \bibnamefont{Archer}},
	\bibinfo{author}{\bibfnamefont{M.~J.} \bibnamefont{Robbins}},
	\bibinfo{author}{\bibfnamefont{H.}~\bibnamefont{Gomez}}, \bibnamefont{and}
	\bibinfo{author}{\bibfnamefont{E.}~\bibnamefont{Knobloch}},
	\bibinfo{journal}{Physical Review E} \textbf{\bibinfo{volume}{87}},
	\bibinfo{pages}{042915} (\bibinfo{year}{2013}).
	
	\bibitem[{\citenamefont{Gavish et~al.}(2017{\natexlab{a}})\citenamefont{Gavish,
			Versano, and Yochelis}}]{gavish2017spatially}
	\bibinfo{author}{\bibfnamefont{N.}~\bibnamefont{Gavish}},
	\bibinfo{author}{\bibfnamefont{I.}~\bibnamefont{Versano}}, \bibnamefont{and}
	\bibinfo{author}{\bibfnamefont{A.}~\bibnamefont{Yochelis}},
	\bibinfo{journal}{SIAM Journal on Applied Dynamical Systems}
	\textbf{\bibinfo{volume}{16}}, \bibinfo{pages}{1946}
	(\bibinfo{year}{2017}{\natexlab{a}}).
	
	\bibitem[{\citenamefont{Belyakov}(1974)}]{Bel:74}
	\bibinfo{author}{\bibfnamefont{L.}~\bibnamefont{Belyakov}},
	\bibinfo{journal}{Mathematical Notes} \textbf{\bibinfo{volume}{15}},
	\bibinfo{pages}{336} (\bibinfo{year}{1974}).
	
	\bibitem[{\citenamefont{Belyakov}(1980)}]{Bel:80}
	\bibinfo{author}{\bibfnamefont{L.}~\bibnamefont{Belyakov}},
	\bibinfo{journal}{Mathematical Notes} \textbf{\bibinfo{volume}{28}},
	\bibinfo{pages}{910} (\bibinfo{year}{1980}).
	
	\bibitem[{\citenamefont{Bre{\~n}a-Medina and
			Champneys}(2014)}]{brena2014subcritical}
	\bibinfo{author}{\bibfnamefont{V.}~\bibnamefont{Bre{\~n}a-Medina}}
	\bibnamefont{and}
	\bibinfo{author}{\bibfnamefont{A.}~\bibnamefont{Champneys}},
	\bibinfo{journal}{Physical Review E} \textbf{\bibinfo{volume}{90}},
	\bibinfo{pages}{032923} (\bibinfo{year}{2014}).
	
	\bibitem[{\citenamefont{Bordiougov and Engel}(2003)}]{BoEn:03}
	\bibinfo{author}{\bibfnamefont{G.}~\bibnamefont{Bordiougov}} \bibnamefont{and}
	\bibinfo{author}{\bibfnamefont{H.}~\bibnamefont{Engel}},
	\bibinfo{journal}{Physical Review Letters} \textbf{\bibinfo{volume}{90}},
	\bibinfo{pages}{148302} (\bibinfo{year}{2003}).
	
	\bibitem[{\citenamefont{Guckenheimer and Holmes}(2013)}]{GuHo:83}
	\bibinfo{author}{\bibfnamefont{J.}~\bibnamefont{Guckenheimer}}
	\bibnamefont{and} \bibinfo{author}{\bibfnamefont{P.}~\bibnamefont{Holmes}},
	\emph{\bibinfo{title}{Nonlinear oscillations, dynamical systems, and
			bifurcations of vector fields}}, vol.~\bibinfo{volume}{42}
	(\bibinfo{publisher}{Springer Science \& Business Media},
	\bibinfo{year}{2013}).
	
	\bibitem[{\citenamefont{Elphick
			et~al.}(1990{\natexlab{a}})\citenamefont{Elphick, Meron, Rinzel, and
			Spiegel}}]{EMRS:90}
	\bibinfo{author}{\bibfnamefont{C.}~\bibnamefont{Elphick}},
	\bibinfo{author}{\bibfnamefont{E.}~\bibnamefont{Meron}},
	\bibinfo{author}{\bibfnamefont{J.}~\bibnamefont{Rinzel}}, \bibnamefont{and}
	\bibinfo{author}{\bibfnamefont{E.}~\bibnamefont{Spiegel}},
	\bibinfo{journal}{Journal of Theoretical Biology}
	\textbf{\bibinfo{volume}{146}}, \bibinfo{pages}{249}
	(\bibinfo{year}{1990}{\natexlab{a}}).
	
	\bibitem[{\citenamefont{Hagberg and Meron}(1998)}]{hagberg1998propagation}
	\bibinfo{author}{\bibfnamefont{A.}~\bibnamefont{Hagberg}} \bibnamefont{and}
	\bibinfo{author}{\bibfnamefont{E.}~\bibnamefont{Meron}},
	\bibinfo{journal}{Physical Review E} \textbf{\bibinfo{volume}{57}},
	\bibinfo{pages}{299} (\bibinfo{year}{1998}).
	
	\bibitem[{\citenamefont{Kiss et~al.}(2004)\citenamefont{Kiss, Merkin, Scott,
			and Simon}}]{kiss2004electric}
	\bibinfo{author}{\bibfnamefont{I.}~\bibnamefont{Kiss}},
	\bibinfo{author}{\bibfnamefont{J.}~\bibnamefont{Merkin}},
	\bibinfo{author}{\bibfnamefont{S.}~\bibnamefont{Scott}}, \bibnamefont{and}
	\bibinfo{author}{\bibfnamefont{P.}~\bibnamefont{Simon}},
	\bibinfo{journal}{The Quarterly Journal of Mechanics and Applied Mathematics}
	\textbf{\bibinfo{volume}{57}}, \bibinfo{pages}{467} (\bibinfo{year}{2004}).
	
	\bibitem[{\citenamefont{Elphick
			et~al.}(1990{\natexlab{b}})\citenamefont{Elphick, Meron, Rinzel, and
			Spiegel}}]{elphick1990impulse}
	\bibinfo{author}{\bibfnamefont{C.}~\bibnamefont{Elphick}},
	\bibinfo{author}{\bibfnamefont{E.}~\bibnamefont{Meron}},
	\bibinfo{author}{\bibfnamefont{J.}~\bibnamefont{Rinzel}}, \bibnamefont{and}
	\bibinfo{author}{\bibfnamefont{E.}~\bibnamefont{Spiegel}},
	\bibinfo{journal}{Journal of Theoretical Biology}
	\textbf{\bibinfo{volume}{146}}, \bibinfo{pages}{249}
	(\bibinfo{year}{1990}{\natexlab{b}}).
	
	\bibitem[{\citenamefont{Or-Guil et~al.}(2000)\citenamefont{Or-Guil, Kevrekidis,
			and B{\"a}r}}]{or2000stable}
	\bibinfo{author}{\bibfnamefont{M.}~\bibnamefont{Or-Guil}},
	\bibinfo{author}{\bibfnamefont{I.~G.} \bibnamefont{Kevrekidis}},
	\bibnamefont{and} \bibinfo{author}{\bibfnamefont{M.}~\bibnamefont{B{\"a}r}},
	\bibinfo{journal}{Physica D: Nonlinear Phenomena}
	\textbf{\bibinfo{volume}{135}}, \bibinfo{pages}{154} (\bibinfo{year}{2000}).
	
	\bibitem[{\citenamefont{R{\"o}der et~al.}(2007)\citenamefont{R{\"o}der,
			Bordyugov, Engel, and Falcke}}]{roder2007wave}
	\bibinfo{author}{\bibfnamefont{G.}~\bibnamefont{R{\"o}der}},
	\bibinfo{author}{\bibfnamefont{G.}~\bibnamefont{Bordyugov}},
	\bibinfo{author}{\bibfnamefont{H.}~\bibnamefont{Engel}}, \bibnamefont{and}
	\bibinfo{author}{\bibfnamefont{M.}~\bibnamefont{Falcke}},
	\bibinfo{journal}{Physical Review E} \textbf{\bibinfo{volume}{75}},
	\bibinfo{pages}{036202} (\bibinfo{year}{2007}).
	
	\bibitem[{\citenamefont{Bonilla and Grahn}(2005)}]{bonilla2005non}
	\bibinfo{author}{\bibfnamefont{L.~L.} \bibnamefont{Bonilla}} \bibnamefont{and}
	\bibinfo{author}{\bibfnamefont{H.~T.} \bibnamefont{Grahn}},
	\bibinfo{journal}{Reports on Progress in Physics}
	\textbf{\bibinfo{volume}{68}}, \bibinfo{pages}{577} (\bibinfo{year}{2005}).
	
	\bibitem[{\citenamefont{D{\"a}hmlow et~al.}(2015)\citenamefont{D{\"a}hmlow,
			Luengviriya, and M{\"u}ller}}]{dahmlow2015electric}
	\bibinfo{author}{\bibfnamefont{P.}~\bibnamefont{D{\"a}hmlow}},
	\bibinfo{author}{\bibfnamefont{C.}~\bibnamefont{Luengviriya}},
	\bibnamefont{and} \bibinfo{author}{\bibfnamefont{S.~C.}
		\bibnamefont{M{\"u}ller}}, in \emph{\bibinfo{booktitle}{Bottom-Up
			Self-Organization in Supramolecular Soft Matter}}
	(\bibinfo{publisher}{Springer}, \bibinfo{year}{2015}), pp.
	\bibinfo{pages}{65--82}.
	
	\bibitem[{\citenamefont{Strubbe and Neyts}(2017)}]{strubbe2017charge}
	\bibinfo{author}{\bibfnamefont{F.}~\bibnamefont{Strubbe}} \bibnamefont{and}
	\bibinfo{author}{\bibfnamefont{K.}~\bibnamefont{Neyts}},
	\bibinfo{journal}{Journal of Physics: Condensed Matter}
	\textbf{\bibinfo{volume}{29}}, \bibinfo{pages}{453003}
	(\bibinfo{year}{2017}).
	
	\bibitem[{\citenamefont{Agladze and De~Kepper}(1992)}]{agladze1992influence}
	\bibinfo{author}{\bibfnamefont{K.}~\bibnamefont{Agladze}} \bibnamefont{and}
	\bibinfo{author}{\bibfnamefont{P.}~\bibnamefont{De~Kepper}},
	\bibinfo{journal}{The Journal of Physical Chemistry}
	\textbf{\bibinfo{volume}{96}}, \bibinfo{pages}{5239} (\bibinfo{year}{1992}).
	
	\bibitem[{\citenamefont{{\v{S}}ev{\v{c}}{\'\i}kov{\'a} and
			M{\"u}ller}(1999)}]{vsevvcikova1999electric}
	\bibinfo{author}{\bibfnamefont{H.}~\bibnamefont{{\v{S}}ev{\v{c}}{\'\i}kov{\'a}}}
	\bibnamefont{and} \bibinfo{author}{\bibfnamefont{S.~C.}
		\bibnamefont{M{\"u}ller}}, \bibinfo{journal}{Physical Review E}
	\textbf{\bibinfo{volume}{60}}, \bibinfo{pages}{532} (\bibinfo{year}{1999}).
	
	\bibitem[{\citenamefont{Sebestikova et~al.}(2005)\citenamefont{Sebestikova,
			Slamova, and Sevcikova}}]{sebestikova2005control}
	\bibinfo{author}{\bibfnamefont{L.}~\bibnamefont{Sebestikova}},
	\bibinfo{author}{\bibfnamefont{E.}~\bibnamefont{Slamova}}, \bibnamefont{and}
	\bibinfo{author}{\bibfnamefont{H.}~\bibnamefont{Sevcikova}},
	\bibinfo{journal}{Biophysical chemistry} \textbf{\bibinfo{volume}{113}},
	\bibinfo{pages}{269} (\bibinfo{year}{2005}).
	
	\bibitem[{\citenamefont{Carballido-Landeira
			et~al.}(2012)\citenamefont{Carballido-Landeira, Taboada, and
			Mu{\~n}uzuri}}]{carballido2012effect}
	\bibinfo{author}{\bibfnamefont{J.}~\bibnamefont{Carballido-Landeira}},
	\bibinfo{author}{\bibfnamefont{P.}~\bibnamefont{Taboada}}, \bibnamefont{and}
	\bibinfo{author}{\bibfnamefont{A.}~\bibnamefont{Mu{\~n}uzuri}},
	\bibinfo{journal}{Soft Matter} \textbf{\bibinfo{volume}{8}},
	\bibinfo{pages}{2945} (\bibinfo{year}{2012}).
	
	\bibitem[{\citenamefont{D{\"a}hmlow and
			M{\"u}ller}(2015)}]{dahmlow2015nonlinear}
	\bibinfo{author}{\bibfnamefont{P.}~\bibnamefont{D{\"a}hmlow}} \bibnamefont{and}
	\bibinfo{author}{\bibfnamefont{S.~C.} \bibnamefont{M{\"u}ller}},
	\bibinfo{journal}{Chaos: An Interdisciplinary Journal of Nonlinear Science}
	\textbf{\bibinfo{volume}{25}}, \bibinfo{pages}{043117}
	(\bibinfo{year}{2015}).
	
	\bibitem[{\citenamefont{Yochelis and
			Sheintuch}(2010{\natexlab{b}})}]{yochelis2010turing}
	\bibinfo{author}{\bibfnamefont{A.}~\bibnamefont{Yochelis}} \bibnamefont{and}
	\bibinfo{author}{\bibfnamefont{M.}~\bibnamefont{Sheintuch}},
	\bibinfo{journal}{Physical Chemistry Chemical Physics}
	\textbf{\bibinfo{volume}{12}}, \bibinfo{pages}{3957}
	(\bibinfo{year}{2010}{\natexlab{b}}).
	
	\bibitem[{\citenamefont{Kuznetsov}(1995)}]{Kuz:95}
	\bibinfo{author}{\bibfnamefont{Y.~A.} \bibnamefont{Kuznetsov}},
	\emph{\bibinfo{title}{Elements of applied bifurcation theory}}
	(\bibinfo{publisher}{Springer-Verlag, NY}, \bibinfo{year}{1995}).
	
	\bibitem[{\citenamefont{Shilnikov et~al.}(1998)\citenamefont{Shilnikov,
			Shilnikov, Turaev, and Chua}}]{SSTC:01}
	\bibinfo{author}{\bibfnamefont{L.~P.} \bibnamefont{Shilnikov}},
	\bibinfo{author}{\bibfnamefont{A.~L.} \bibnamefont{Shilnikov}},
	\bibinfo{author}{\bibfnamefont{D.~V.} \bibnamefont{Turaev}},
	\bibnamefont{and} \bibinfo{author}{\bibfnamefont{L.~O.} \bibnamefont{Chua}},
	\emph{\bibinfo{title}{Methods Of Qualitative Theory In Nonlinear Dynamics:
			Part II}} (\bibinfo{publisher}{World Scientific}, \bibinfo{year}{1998}).
	
	\bibitem[{\citenamefont{Chomaz et~al.}(1999)\citenamefont{Chomaz, Couairon, and
			Julien}}]{chomaz1999absolute}
	\bibinfo{author}{\bibfnamefont{J.-M.} \bibnamefont{Chomaz}},
	\bibinfo{author}{\bibfnamefont{A.}~\bibnamefont{Couairon}}, \bibnamefont{and}
	\bibinfo{author}{\bibfnamefont{S.}~\bibnamefont{Julien}},
	\bibinfo{journal}{Physics of Fluids} \textbf{\bibinfo{volume}{11}},
	\bibinfo{pages}{3369} (\bibinfo{year}{1999}).
	
	\bibitem[{\citenamefont{Kuznetsov and Hooman}(2008)}]{KuHo:08}
	\bibinfo{author}{\bibfnamefont{A.}~\bibnamefont{Kuznetsov}} \bibnamefont{and}
	\bibinfo{author}{\bibfnamefont{K.}~\bibnamefont{Hooman}},
	\bibinfo{journal}{International Journal of Heat and Mass Transfer}
	\textbf{\bibinfo{volume}{51}}, \bibinfo{pages}{5695} (\bibinfo{year}{2008}).
	
	\bibitem[{\citenamefont{Nagahara et~al.}(2009)\citenamefont{Nagahara, Ma,
			Takenaka, Kageyama, and Yoshikawa}}]{NMTKY:09}
	\bibinfo{author}{\bibfnamefont{H.}~\bibnamefont{Nagahara}},
	\bibinfo{author}{\bibfnamefont{Y.}~\bibnamefont{Ma}},
	\bibinfo{author}{\bibfnamefont{Y.}~\bibnamefont{Takenaka}},
	\bibinfo{author}{\bibfnamefont{R.}~\bibnamefont{Kageyama}}, \bibnamefont{and}
	\bibinfo{author}{\bibfnamefont{K.}~\bibnamefont{Yoshikawa}},
	\bibinfo{journal}{Physical Review E} \textbf{\bibinfo{volume}{80}},
	\bibinfo{pages}{021906} (\bibinfo{year}{2009}).
	
	\bibitem[{\citenamefont{Sherratt}(2011)}]{sherratt2011pattern}
	\bibinfo{author}{\bibfnamefont{J.~A.} \bibnamefont{Sherratt}}, in
	\emph{\bibinfo{booktitle}{Proc. R. Soc. A}} (\bibinfo{organization}{The Royal
		Society}, \bibinfo{year}{2011}), vol. \bibinfo{volume}{467}, pp.
	\bibinfo{pages}{3272--3294}.
	
	\bibitem[{\citenamefont{Siero et~al.}(2015)\citenamefont{Siero, Doelman,
			Eppinga, Rademacher, Rietkerk, and Siteur}}]{siero2015striped}
	\bibinfo{author}{\bibfnamefont{E.}~\bibnamefont{Siero}},
	\bibinfo{author}{\bibfnamefont{A.}~\bibnamefont{Doelman}},
	\bibinfo{author}{\bibfnamefont{M.}~\bibnamefont{Eppinga}},
	\bibinfo{author}{\bibfnamefont{J.~D.} \bibnamefont{Rademacher}},
	\bibinfo{author}{\bibfnamefont{M.}~\bibnamefont{Rietkerk}}, \bibnamefont{and}
	\bibinfo{author}{\bibfnamefont{K.}~\bibnamefont{Siteur}},
	\bibinfo{journal}{Chaos} \textbf{\bibinfo{volume}{25}},
	\bibinfo{pages}{036411} (\bibinfo{year}{2015}).
	
	\bibitem[{\citenamefont{Berenstein}(2012)}]{berenstein2012distinguishing}
	\bibinfo{author}{\bibfnamefont{I.}~\bibnamefont{Berenstein}},
	\bibinfo{journal}{Chaos} \textbf{\bibinfo{volume}{22}},
	\bibinfo{pages}{043109} (\bibinfo{year}{2012}).
	
	\bibitem[{\citenamefont{Ghosh et~al.}(2016)\citenamefont{Ghosh, Paul, and
			Ray}}]{ghosh2016differential}
	\bibinfo{author}{\bibfnamefont{S.}~\bibnamefont{Ghosh}},
	\bibinfo{author}{\bibfnamefont{S.}~\bibnamefont{Paul}}, \bibnamefont{and}
	\bibinfo{author}{\bibfnamefont{D.~S.} \bibnamefont{Ray}},
	\bibinfo{journal}{Physical Review E} \textbf{\bibinfo{volume}{94}},
	\bibinfo{pages}{042223} (\bibinfo{year}{2016}).
	
	\bibitem[{\citenamefont{Yochelis et~al.}(2016)\citenamefont{Yochelis, Bar-On,
			and Gov}}]{yochelis2016reaction}
	\bibinfo{author}{\bibfnamefont{A.}~\bibnamefont{Yochelis}},
	\bibinfo{author}{\bibfnamefont{T.}~\bibnamefont{Bar-On}}, \bibnamefont{and}
	\bibinfo{author}{\bibfnamefont{N.~S.} \bibnamefont{Gov}},
	\bibinfo{journal}{Physica D} \textbf{\bibinfo{volume}{318-319}},
	\bibinfo{pages}{84} (\bibinfo{year}{2016}).
	
	\bibitem[{\citenamefont{Holzer and Popovic}(2017)}]{holzer2017wavetrain}
	\bibinfo{author}{\bibfnamefont{M.}~\bibnamefont{Holzer}} \bibnamefont{and}
	\bibinfo{author}{\bibfnamefont{N.}~\bibnamefont{Popovic}},
	\bibinfo{journal}{SIAM Journal on Applied Dynamical Systems}
	\textbf{\bibinfo{volume}{16}}, \bibinfo{pages}{431} (\bibinfo{year}{2017}).
	
	\bibitem[{\citenamefont{Vidal-Henriquez
			et~al.}(2017)\citenamefont{Vidal-Henriquez, Zykov, Bodenschatz, and
			Gholami}}]{vidal2017convective}
	\bibinfo{author}{\bibfnamefont{E.}~\bibnamefont{Vidal-Henriquez}},
	\bibinfo{author}{\bibfnamefont{V.}~\bibnamefont{Zykov}},
	\bibinfo{author}{\bibfnamefont{E.}~\bibnamefont{Bodenschatz}},
	\bibnamefont{and} \bibinfo{author}{\bibfnamefont{A.}~\bibnamefont{Gholami}},
	\bibinfo{journal}{Chaos} \textbf{\bibinfo{volume}{27}},
	\bibinfo{pages}{103110} (\bibinfo{year}{2017}).
	
	\bibitem[{\citenamefont{Yochelis
			et~al.}(2015{\natexlab{b}})\citenamefont{Yochelis, Ebrahim, Millis, Cui,
			Kachar, Naoz, and Gov}}]{yochelis2015self}
	\bibinfo{author}{\bibfnamefont{A.}~\bibnamefont{Yochelis}},
	\bibinfo{author}{\bibfnamefont{S.}~\bibnamefont{Ebrahim}},
	\bibinfo{author}{\bibfnamefont{B.}~\bibnamefont{Millis}},
	\bibinfo{author}{\bibfnamefont{R.}~\bibnamefont{Cui}},
	\bibinfo{author}{\bibfnamefont{B.}~\bibnamefont{Kachar}},
	\bibinfo{author}{\bibfnamefont{M.}~\bibnamefont{Naoz}}, \bibnamefont{and}
	\bibinfo{author}{\bibfnamefont{N.}~\bibnamefont{Gov}},
	\bibinfo{journal}{Scientific Reports} \textbf{\bibinfo{volume}{5}},
	\bibinfo{pages}{13521} (\bibinfo{year}{2015}{\natexlab{b}}).
	
	\bibitem[{\citenamefont{Gavish et~al.}(2017{\natexlab{b}})\citenamefont{Gavish,
			Elad, and Yochelis}}]{gavish2017solvent}
	\bibinfo{author}{\bibfnamefont{N.}~\bibnamefont{Gavish}},
	\bibinfo{author}{\bibfnamefont{D.}~\bibnamefont{Elad}}, \bibnamefont{and}
	\bibinfo{author}{\bibfnamefont{A.}~\bibnamefont{Yochelis}},
	\bibinfo{journal}{Journal of Physical Chemistry Letters}
	\textbf{\bibinfo{volume}{9}}, \bibinfo{pages}{36}
	(\bibinfo{year}{2017}{\natexlab{b}}).
	
	\bibitem[{\citenamefont{Siebert et~al.}(2014)\citenamefont{Siebert, Alonso,
			B{\"a}r, and Sch{\"o}ll}}]{siebert2014dynamics}
	\bibinfo{author}{\bibfnamefont{J.}~\bibnamefont{Siebert}},
	\bibinfo{author}{\bibfnamefont{S.}~\bibnamefont{Alonso}},
	\bibinfo{author}{\bibfnamefont{M.}~\bibnamefont{B{\"a}r}}, \bibnamefont{and}
	\bibinfo{author}{\bibfnamefont{E.}~\bibnamefont{Sch{\"o}ll}},
	\bibinfo{journal}{Physical Review E} \textbf{\bibinfo{volume}{89}},
	\bibinfo{pages}{052909} (\bibinfo{year}{2014}).
	
	\bibitem[{\citenamefont{Brooks and Bressloff}(2016)}]{brooks2016mechanism}
	\bibinfo{author}{\bibfnamefont{H.~A.} \bibnamefont{Brooks}} \bibnamefont{and}
	\bibinfo{author}{\bibfnamefont{P.~C.} \bibnamefont{Bressloff}},
	\bibinfo{journal}{SIAM Journal on Applied Dynamical Systems}
	\textbf{\bibinfo{volume}{15}}, \bibinfo{pages}{1823} (\bibinfo{year}{2016}).
	
	\bibitem[{\citenamefont{Zmurchok et~al.}(2017)\citenamefont{Zmurchok, Small,
			Ward, and Edelstein-Keshet}}]{zmurchok2017application}
	\bibinfo{author}{\bibfnamefont{C.}~\bibnamefont{Zmurchok}},
	\bibinfo{author}{\bibfnamefont{T.}~\bibnamefont{Small}},
	\bibinfo{author}{\bibfnamefont{M.~J.} \bibnamefont{Ward}}, \bibnamefont{and}
	\bibinfo{author}{\bibfnamefont{L.}~\bibnamefont{Edelstein-Keshet}},
	\bibinfo{journal}{Bulletin of Mathematical Biology}
	\textbf{\bibinfo{volume}{79}}, \bibinfo{pages}{1923} (\bibinfo{year}{2017}).
	
	\bibitem[{\citenamefont{Altimari et~al.}(2012)\citenamefont{Altimari, Mancusi,
			and Crescitelli}}]{altimari2012formation}
	\bibinfo{author}{\bibfnamefont{P.}~\bibnamefont{Altimari}},
	\bibinfo{author}{\bibfnamefont{E.}~\bibnamefont{Mancusi}}, \bibnamefont{and}
	\bibinfo{author}{\bibfnamefont{S.}~\bibnamefont{Crescitelli}},
	\bibinfo{journal}{Industrial \& Engineering Chemistry Research}
	\textbf{\bibinfo{volume}{51}}, \bibinfo{pages}{9609} (\bibinfo{year}{2012}).
	
	\bibitem[{\citenamefont{Berenstein and Beta}(2012)}]{berenstein2012flow}
	\bibinfo{author}{\bibfnamefont{I.}~\bibnamefont{Berenstein}} \bibnamefont{and}
	\bibinfo{author}{\bibfnamefont{C.}~\bibnamefont{Beta}},
	\bibinfo{journal}{Physical Review E} \textbf{\bibinfo{volume}{86}},
	\bibinfo{pages}{056205} (\bibinfo{year}{2012}).
	
\end{thebibliography}

\end{document}